\documentclass[12pt]{iopart}
\usepackage{iopams}
\usepackage{enumitem}
\usepackage{graphicx}
\usepackage{cite}
\usepackage{braket}
\newcommand{\vn}{\mathbf{n}}
\newcommand{\aver}[1]{\langle #1 \rangle}
\newcommand{\thetaa}{\theta_{\textsc{a}}}
\newcommand{\thetaat}{\theta_{\textsc{a}}}
\newcommand{\thetab}{\theta_{\textsc{b}}}
\newcommand{\thetabt}{\theta_{\textsc{b}}}
\graphicspath{{./figures/}}
\newcommand{\af}{\textsc{af}}

\newcommand{\neweps}{\varepsilon}
\newcommand{\epsarg}[1]{\varepsilon_{\small\textsc{#1}}}

\begin{document}

\title[Tomographic indicators at avoided crossings]{Signatures of avoided energy-level crossings in entanglement indicators obtained from quantum tomograms}

\author{B. Sharmila$^{1}$, S. Lakshmibala$^{1}$ and V. Balakrishnan$^{1}$}

\address{$^{1}$ Department of Physics, Indian Institute of Technology Madras, Chennai 600036, India.}
\ead{sharmilab@physics.iitm.ac.in}
\vspace{10pt}
\begin{indented}
\item[]\today
\end{indented}

\begin{abstract}
Extensive theoretical and experimental investigations on multipartite systems close to an avoided energy-level crossing reveal interesting
features  such as the extremisation  of entanglement. Conventionally, the  estimation of entanglement directly from experimental observation involves either one of two approaches:
Uncertainty-relation-based estimation that captures the linear correlation between relevant observables, or rigorous but error-prone quantum state reconstruction on tomograms obtained from homodyne measurements. We investigate the
behaviour,  close to avoided crossings,  
of entanglement indicators that can be
calculated directly from a numerically-generated tomogram. The systems we study are
two generic bipartite continuous-variable
systems: a Bose-Einstein condensate trapped in a double-well potential,  and a multi-level atom interacting with a radiation field. We also
consider a multipartite hybrid quantum system of superconducting qubits interacting with microwave photons.
We carry out a quantitative comparison of the indicators with a standard measure of entanglement, the subsystem von Neumann entropy (SVNE). It is shown that the  indicators that capture the nonlinear correlation between relevant subsystem observables are in excellent agreement with the SVNE.
\end{abstract}
\vspace{2pc}

\submitto{\JPB}

\maketitle
%
%
\section{\label{sec:intro}Introduction}

The measurement of any  observable in  a quantum mechanical system yields a histogram of the state of the system  in the basis of that observable. In particular, in the context of quantum optics, measurements of a judiciously chosen  quorum of field observables yield a set of histograms (the optical tomogram) from which the density matrix is reconstructed.
 The latter is needed in standard  procedures  for estimating the extent of entanglement between the  subsystems of  bipartite or multipartite systems. A standard   measure of the entanglement between the two subsystems of a bipartite system is the subsystem von Neumann entropy
$\xi_{\textsc{svne}}=
-\Tr\,(\rho \,\log_{2} \,\rho)$,  where $\rho$ is the density matrix of either one of the subsystems~\cite{niel}. In the case of continuous-variable (CV) quantum systems, an infinite set of histograms is required,  in principle,  in order to obtain complete information about the density matrix. In practice, however,  only a finite set of histograms (corresponding to measurement of a finite set of observables) can be obtained. As state reconstruction from tomograms~\cite{QSRecon} typically involves error-prone statistical techniques such as maximum likelihood estimates,  it is preferable to assess the extent of entanglement directly  from tomograms, circumventing detailed state reconstruction.
 
In earlier work ~\cite{sharmila} we  have proposed a tomographic   entanglement indicator 
$\xi_{\textsc{tei}}$  based on mutual information, that is obtained directly from the relevant tomograms. We have  tested its efficacy in bipartite CV systems evolving unitarily  under nonlinear Hamiltonians, by comparing it both with $\xi_{\textsc{svne}}$ and with an entanglement indicator $\xi_{\textsc{ipr}}$ based on inverse participation ratios~\cite{sharmila2}. (The participation ratio is a measure of delocalization in a given basis.) It has been  shown that $\xi_{\textsc{tei}}$ and $\xi_{\textsc{ipr}}$ capture the gross features of entanglement dynamics. A time-series analysis of the difference $|\xi_{\textsc{tei}}-\xi_{\textsc{svne}}|$  was used to quantify the deviation   of $\xi_{\textsc{tei}}$ from $ \xi_{\textsc{svne}}$ as a function of time.  Although this is sensitive to the specific choice of initial state and the strength of the nonlinearity in the Hamiltonian, it has been shown that $\xi_{\textsc{tei}}$ is a reasonably good indicator of entanglement in general. In  multipartite hybrid quantum (HQ) systems comprising two-level atoms interacting with radiation fields, too, $\xi_{\textsc{tei}}$ turns out~\cite{sharmila3} to be  a good estimator of  quantum correlations in both the field and the atomic subsystems. (In the atom sector,  $\xi_{\textsc{tei}}$ is extracted directly from the corresponding qubit tomograms).

Several indicators of correlations  between different parts of a {\em classical} system  have been used extensively in various applications such  as automated image processing. These correlators are obtained from classical tomograms. Their definitions, however,  are not intrinsically classical in nature,  and it is worth examining their applicability in quantum contexts. Since correlations are inherently present in entangled states of quantum systems, a natural question that arises is whether the  performance of entanglement quantifiers obtained from these correlators is comparable to that of standard indicators such as  $\xi_{\textsc{svne}}$.  

We examine quantum systems where the spacings between energy levels change significantly with changes in the parameters, with two or more levels moving close to each other for specific values of the parameters and then moving away as these values change. We will henceforth refer to this feature as avoided energy-level crossing. Extensive studies~\cite{AClinkPT,ent_change,ent_max} have established that entanglement (as measured by standard indicators such as $\xi_{\textsc{svne}}$) is generically at an extremum  at an avoided crossing. Typically, the energy spectrum and the spacing between the energy levels depend on the strengths of the nonlinearity and the coupling between subsystems. With changes in the values of  these  parameters, the spacing between adjacent  levels can decrease, and even tend to zero, resulting in an energy-level crossing. According to the von Neumann-Wigner no-crossing theorem, energy levels within a multiplet generically avoid crossing, provided only one of the parameters is varied in the Hamiltonian governing the system. In this paper, we  investigate how effectively some of these entanglement indicators mimic the behaviour of $\xi_{\textsc{svne}}$ close to avoided crossings.

Energy-level  crossings display other interesting features. Since they affect the level spacings and their probability distribution ~\cite{BGSconj},  they  are
also important from the point of view of non-integrability and quantum chaos (see, for instance,~\cite{haake}). In addition, avoided crossings point to phase transitions which trigger a change in the quantum correlations in the system ~\cite{cejnar,heiss,AClinkPT,ent_change}. This aspect has been investigated extensively both  theoretically and in experiments~\cite{ACtheor1,ACexpt1,ACexpt2,ACexpt3}.

We examine  two experimentally relevant  bipartite CV systems: a Bose-Einstein condensate (BEC) in a double-well trap~\cite{sanz},  and a multilevel atom interacting with a radiation field~\cite{agarwalpuri}. We also investigate a multipartite HQ system~\cite{QuantMeta2,QuantMeta} that is effectively described by the Tavis-Cummings model
~\cite{tavis}. The rest of this paper is organized as follows: In the next section  we introduce the entanglement indicators  to be employed.  In \sref{sec:work}, we investigate how these indicators behave close to avoided crossings in the two  bipartite CV models mentioned above.  In \sref{sec:work2}, we extend our analysis to the multipartite HQ model.
Concluding  remarks are made in  \sref{sec:conc}.

\section{\label{sec:review}Entanglement indicators from tomograms}

We first consider generic CV systems. 
A typical example of a bipartite CV system  is two coupled oscillators (equivalently, a 
single-mode  radiation field interacting with a multilevel atom modelled as an oscillator).    The  tomogram is obtained from  the quorum of observables that contain complete information  about the state. These  observables are represented by  the rotated quadrature operators
\begin{equation}
\mathbb{X}_{\thetaa} = (a \,e^{-i \thetaat} + a^{\dagger} \,e^{i \thetaat})/\sqrt{2},\;\;
 \mathbb{X}_{\thetab} = (b e^{-i \thetabt} + b^{\dagger} e^{i \thetabt})/\sqrt{2}.
\label{eqn:rotQuadOps}
\end{equation}
Here $0 \leqslant  \thetaa, \thetab < \pi$, and $(a, a^{\dagger})$ [respectively, $(b, b^{\dagger})$] are the oscillator annihilation and creation operators corresponding to the two subsystems  A and B. The bipartite tomogram is given by 
\begin{equation}
w(X_{\thetaat},\thetaa;  X_{\thetabt},\thetab) =
\bra{X_{\thetaat},\thetaa;X_{\thetabt},\thetab} \rho_{\textsc{ab}} \ket{X_{\thetaat},\thetaa;X_{\thetabt},\thetab}, 
\label{eqn:tomoDefn}
\end{equation}
where $\rho_{\textsc{ab}}$ denotes the bipartite density matrix. Here $\mathbb{X}_{\theta_{i}}\ket{X_{\theta_{i}},\theta_{i}}=X_{\theta_{i}}\ket{X_{\theta_{i}},\theta_{i}}$ ($i$=A,B),  and  the product  basis state $\ket{X_{\thetaat},\thetaa} \otimes \ket{X_{\thetabt},\thetab}$ is written as 
$\ket{X_{\thetaat},\thetaa;X_{\thetabt},\thetab}$. The normalization condition is given by 
\begin{equation}
\int_{-\infty}^{\infty}\! \rmd X_{\thetaat} \int_{-\infty}^{\infty} \!\rmd X_{\thetabt} w(X_{\thetaat},\thetaa;X_{\thetabt},\thetab) = 1
\label{eqn:tomoNorm}
\end{equation}
for each $\thetaa$ and $\thetab$. 
The reduced tomogram for subsystem A is   
\begin{eqnarray}
\nonumber w_{\textsc{a}}(X_{\thetaat},\thetaa) & = \int_{-\infty}^{\infty} \rmd X_{\thetabt} w(X_{\thetaat},\thetaa;X_{\thetabt},\thetab)\\[4pt]
& = \bra{X_{\thetaat},\thetaa} \rho_{\textsc{a}} \ket{X_{\thetaat},\thetaa},
\label{eqn:RedTomo}
\end{eqnarray}
where $\rho_{\textsc{a}}=\mathrm{Tr_{B}}\,(\rho_{\textsc{ab}})$ is the  corresponding reduced density matrix. A similar definition holds for  subsystem B.  In order to estimate the degree of  correlation between the subsystems, we use the following tomographic entropies.
The bipartite tomographic entropy is given by 
\begin{eqnarray}
\nonumber S(\thetaa,\thetab) = - & \int_{-\infty}^{\infty} \, \rmd X_{\thetaat} \int_{-\infty}^{\infty} \, \rmd X_{\thetabt} w(X_{\thetaat},\thetaa;X_{\thetabt},\thetab)\,\times \\[4pt] 
 & \log_{2} \, w(X_{\thetaat},\thetaa;X_{\thetabt},\thetab).
\label{eqn:2modeEntropy}
\end{eqnarray}
The subsystem tomographic entropy is
\begin{eqnarray}
S(\theta_{i}) = - &\int_{-\infty}^{\infty} \!\rmd X_{\theta_{i}} w_{i}(X_{\theta_{i}},\theta_{i}) \log_{2} \,[w_{i}(X_{\theta_{i}},\theta_{i})] \;\; (i= \textsc{A,B}).
\label{eqn:1modeEntropy}
\end{eqnarray}
Some of the correlators that we examine in this paper  are obtained from a  section of the tomogram
 corresponding to specific values of $\thetaat$ and $\thetabt$. The efficacy  of such a correlator  
 as a measure  of entanglement is therefore sensitive to the choice of  the tomographic section. We now define these  correlators, and the corresponding entanglement indicators.  

The mutual information 
$\epsarg{tei}(\thetaa,\thetab)$  which we get from the tomogram of a quantum system 
can carry  signatures of entanglement. This  
quantity is expressed in terms of  the  tomographic entropies defined above  as
\begin{equation}
\epsarg{tei}(\thetaa,\thetab)=S(\thetaa) + S(\thetab) - S(\thetaa,\thetab).
\label{eqn:epsTEI}
\end{equation}
Indicators based on the inverse participation ratio (IPR) are also found to be good candidates for estimating  the extent of entanglement ~\cite{ViolaBrown,sharmila2}.  The IPR corresponding to a bipartite system in the basis of the rotated quadrature operators  is defined as
\begin{equation}
\eta_{\textsc{ab}}(\thetaa,\thetab) = \int_{-\infty}^{\infty} \!\rmd X_{\thetaat} \int_{-\infty}^{\infty} \!\rmd X_{\thetabt} [w(X_{\thetaat},\thetaa;X_{\thetabt},\thetab)]^{2}.
\label{eqn:IPRab}
\end{equation}
The IPR for each subsystem is given by 
\begin{equation}
\eta_{i}(\theta_{i}) = \int_{-\infty}^{\infty} \!\rmd X_{\theta_{i}} [w_{i}(X_{\theta_{i}},\theta_{i})]^{2} \;\; 
(i= \textsc{A,B}).
\label{eqn:IPRi}
\end{equation}
The entanglement indicator  in this case is given by 
\begin{equation}
\epsarg{ipr}(\thetaa,\thetab)=1+\eta_{\textsc{ab}}(\thetaa,\thetab)-\eta_{\textsc{a}}(\thetaa)-\eta_{\textsc{b}}(\thetab). 
\label{eqn:epsIPR}
\end{equation}  

Apart from these, we have examined two  
other correlators which are familiar in the context of classical tomograms. The first of these is the Pearson correlation coefficient~\cite{SmithMIcorr}   between two random variables $X$ and $Y$, given by 
\begin{equation}
\nonumber 
\textsc{PCC}(X,Y)
 = \frac{\mathrm{Cov}(X,Y)}{\sigma_{\textsc{x}} \sigma_{\textsc{y}}}.
\label{eqn:PCC}
\end{equation}
Here $\sigma_{\textsc{x}}, \sigma_{\textsc{y}}$ are  the standard deviations of $X$ and $Y$ respectively, and  
$\mathrm{Cov}(X,Y)$ is their covariance.
Of direct relevance to us is 
$\textsc{PCC}(X_{\thetaat},X_{\thetabt})$ 
calculated  for  fixed values of $\thetaa$ and $\thetab$.
Since the quantifier  of entanglement between two subsystems must be non-negative, a simple definition of the entanglement indicator in this case would be 
\begin{equation}
\epsarg{pcc}(\thetaa,\thetab)=
\vert \textsc{PCC}(X_{\thetaat},X_{\thetabt})\vert.
\label{eqn:epsPCC}
\end{equation}
This indicator captures  the effect of linear correlations.  Our motivation for assessing this indicator arises from the fact that,   in recent experiments on generating and testing the extent of entanglement in CV systems, the variances of suitably chosen conjugate observables and the corresponding standard quantum limit alone are used ~\cite{ExptCVentang}. 
We reiterate that these merely capture the extent of linear correlations between two states. 

The second indicator (to be denoted by 
$\epsarg{bd}$) that we introduce and use is arrived at as follows. 
In probability theory, 
the mutual information~\cite{ITCoverThomas}  
 between two continuous random variables 
$X$ and $Y$ 
can be expressed in terms of the Kullback-Leibler divergence
$D_{\textsc{kl}}$~\cite{kullback} 
between their joint probability density  
$p_{XY}(x,y)$ 
and the product of the corresponding  marginal 
densities  $p_{X}(x)= \int  p_{XY}(x,y) dy$ 
and $p_{Y}(y) = \int p_{XY}(x,y) dx$,    
as~\cite{MI_KL_link} 
\begin{equation}
D_{\textsc{kl}}[p_{XY}\!:\!p_{X}p_{Y}] 
= \int \!dx \int\!dy \,p_{XY}(x,y)\,\log_{2} \,
\frac{p_{XY}(x,y)}{p_{X}(x) p_{Y}(y)},
\label{eqn:KLmutualinfo}
\end{equation}
The quantity  $\epsarg{tei}(\thetaa,\thetab)$ 
defined in  Eq. \eref{eqn:epsTEI} 
is precisely the mutual information in the case of 
optical tomograms (which are continuous probability distributions): 
\begin{equation}
\epsarg{tei}(\thetaa,\thetab)= 
D_{\textsc{kl}}
\big[w(X_{\thetaat},\thetaa;X_{\thetabt},\thetab)\!:\! w_{\textsc{a}}(X_{\thetaat},\theta) w_{\textsc{b}}(X_{\thetabt},\thetab)\big]. 
\label{eqn:KL_MI_link}
\end{equation}
 A simpler  alternative for our purposes is 
provided by the Bhattacharyya distance
$D_{\textsc{b}}$~\cite{KL_BD_link} between $p_{XY}$ and 
$p_{X}p_{Y}$, defined as  
\begin{equation}
D_{\textsc{b}}[p_{XY}\!:\!p_{X}p_{Y}] =  
 -\log_{2}\,\Big\{\int \!dx\int \!dy \,\big[p_{xy}(x,y) 
p_{X}(x) p_{Y}(y)\big]^{1/2}\Big\}. 
\label{eqn:Bhattmutualinfo}
\end{equation}
Using Jensen's inequality, it is easily shown that 
$D_{\textsc{b}} \leqslant \frac{1}{2} D_{\textsc{kl}}$. 
$D_{\textsc{b}}$ thus gives us an approximate estimate (that is an underestimate) 
of the mutual information. Based on this quantity,  
 we have 
 an entanglement indicator that is the 
 analogue of Eq. (\ref{eqn:KL_MI_link}), namely, 
\begin{eqnarray}
\epsarg{bd}(\thetaa,\thetab)=& D_{\textsc{b}}[w(X_{\thetaat},\thetaa;X_{\thetabt},\thetab)\!:\! w_{\textsc{a}}(X_{\thetaat},\theta) w_{\textsc{b}}(X_{\thetabt},\thetab)].
\label{eqn:epsBD}
\end{eqnarray}

The dependence on 
$\thetaa$ and $\thetab$ of each of the foregoing entanglement 
 indicators $\varepsilon$ is removed by  
  averaging over a representative set of values of  
  those variables. We denote the 
corresponding averaged value by $\xi$. 
 In the context  of 
 bipartite CV models,     
 we have shown in earlier work~\cite{sharmila,sharmila2}  that  averaging 
 $\epsarg{tei}(\thetaa,\thetab)$ 
over  $25$ different values of 
 $(\thetaa,\thetab)$ selected at equal intervals in the range $[0,\pi)$ yields a reliable  
entanglement indicator  $\xi_{\textsc{tei}}$.
 A similar averaging of each of the quantities 
$\epsarg{ipr}, \epsarg{pcc}$ and 
$\epsarg{bd}$ yields 
  $\xi_{\textsc{ipr}}, \xi_{\textsc{pcc}}$ and   
 $\xi_{\textsc{bd}}$, respectively. 

Next, we turn  to  hybrid systems of field-atom interactions. For a two-level atom with ground state $\ket{g}$ and excited state $\ket{e}$, the quorum of observables  is ~\cite{thew} 
\begin{eqnarray}
\label{eqn:atomops}
\nonumber \sigma_{x}=
\textstyle{\frac{1}{2}} (\ket{e}\bra{g}&+\ket{g}\bra{e}), 
\;\sigma_{y}=\textstyle{\frac{1}{2}} i(\ket{g}\bra{e}-\ket{e}\bra{g}), \\
& \sigma_{z}=\textstyle{\frac{1}{2}} (\ket{e}\bra{e}-\ket{g}\bra{g}), 
\end{eqnarray}
where  $\sigma_{i}$ is a Pauli matrix.
Let $\sigma_{z}\ket{m}=m\ket{m}$. Then $U(\vartheta,\varphi) \ket{m} =  \ket{\vartheta,\varphi,m}$, where $U(\vartheta,\varphi)$ 
 is a general SU(2) transformation parametrized by $(\vartheta, \varphi)$. Denoting $(\vartheta, \varphi)$ by the unit vector $\vn$, the  qubit tomogram is  given by 
\begin{equation}
\label{eqn:spintomogram}
w(\vn,m)=\bra{\vn,m}\rho_{\textsc{s}}\ket{\vn,m}
\end{equation}
where $\rho_{\textsc{s}}$ is the qubit density matrix. Corresponding to each value of $\vn$ there exists a complete basis set.  The atomic tomograms 
are  obtained from these,  and the corresponding 
entanglement properties are quantified using  
appropriate adaptations of the indicators described above.
The  extension of the foregoing to the multipartite 
case is straightforward ~\cite{ibort}, and the tomograms 
obtained can be examined on similar lines.  

\section{\label{sec:work}Avoided energy-level crossings in bipartite CV models}

\subsection{The double-well BEC model}

The effective Hamiltonian for the system and its diagonalisation are as follows~\cite{sanz}. 
Setting $\hbar = 1$,
\begin{equation}
H_{\textsc{bec}} =\omega_{0} N_{\mathrm{tot}} + \omega_{1} (a^{\dagger} a - b^{\dagger} b) + U  N_{\mathrm{tot}}^{2} - \lambda (a^{\dagger} b + a b^{\dagger}).
\label{eqn:HBEC}
\end{equation}
Here, $(a,a^{\dagger})$ and $(b,b^{\dagger})$ are the  respective boson annihilation and creation operators of the atoms in wells A and B (the two subsystems),  
and $N_{\mathrm{tot}} 
= (a^{\dagger} a + b^{\dagger} b)$.
 $U$ is the strength of  nonlinear interactions between atoms within each well, and also between  the two wells. $U>0$, ensuring that the energy spectrum is bounded from below. $\lambda$ is the linear interaction strength, while $\omega_{1}$ is the strength of
 the population imbalance between the two wells.   
 The Hamiltonian is diagonalised by the unitary 
 transformation 
$V= e^{\kappa (a^{\dagger} b - b^{\dagger} a)/2}$  where $\kappa
=\tan^{-1} (\lambda/\omega_{1})$, to yield 
\begin{equation} 
 V^{\dagger} H_{\textsc{bec}} V = 
 \widetilde{H}_{\textsc{bec}}
= \omega_{0} N_{\mathrm{tot}} + \lambda_{1} (a^{\dagger} a - b^{\dagger} b) + 
U  N_{\mathrm{tot}}^{2},
\label{eqn:VrotatedH}
\end{equation}
with $\lambda_{1}= (\lambda^{2} + 
\omega_{1}^{2})^{1/2}$.  
$\widetilde{H}_{\textsc{bec}}$ and $N_{\mathrm{tot}}$ 
commute with each other.
Their common eigenstates   are 
the  product states $\ket{k} \otimes \ket{N-k} \equiv 
\ket{k, N-k}$. Here  
$N = 0,1,2,\dots$ 
is the eigenvalue of 
$N_{\mathrm{tot}}$, 
and $\ket{k}$ is a 
 boson number state, with $k$ running from 
 $0$ to $N$ for a given $N$.
  The eigenstates and 
 eigenvalues of 
$H_{\textsc{bec}}$ are given by
\begin{equation}
\ket{\psi_{N,k}} = V \ket{k, N-k}
\label{eqn:EigvecBEC}
\end{equation}
and
\begin{equation}
E(N,k)= \omega_{0} N + \lambda_{1} (2 k - N) + U  N^{2}.
\label{eqn:EigvalBEC}
\end{equation}
For numerical analysis we set $\omega_{0} = 1, 
 U =1$.

\begin{figure}
\includegraphics[width=0.45\textwidth]{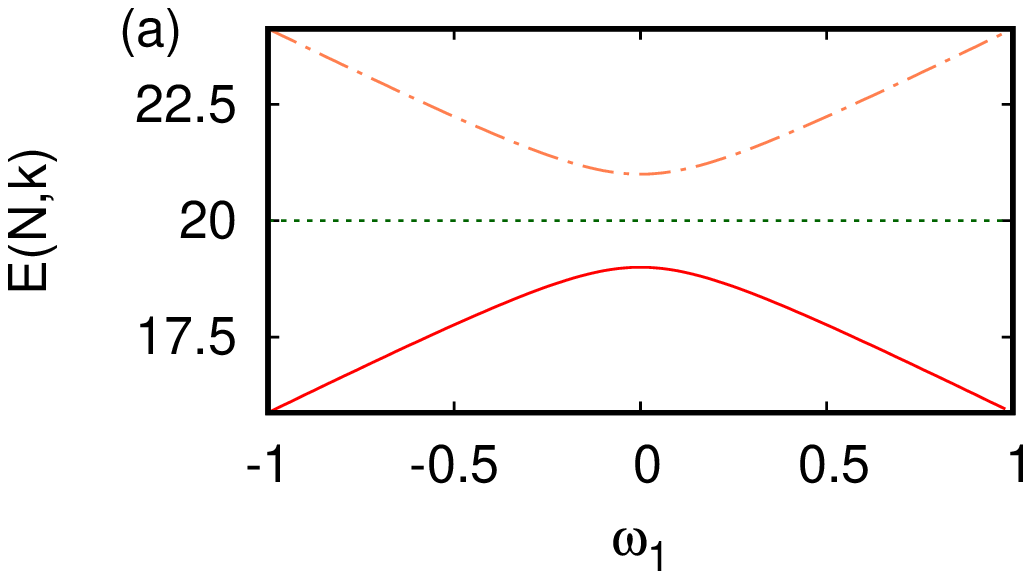}
\includegraphics[width=0.45\textwidth]{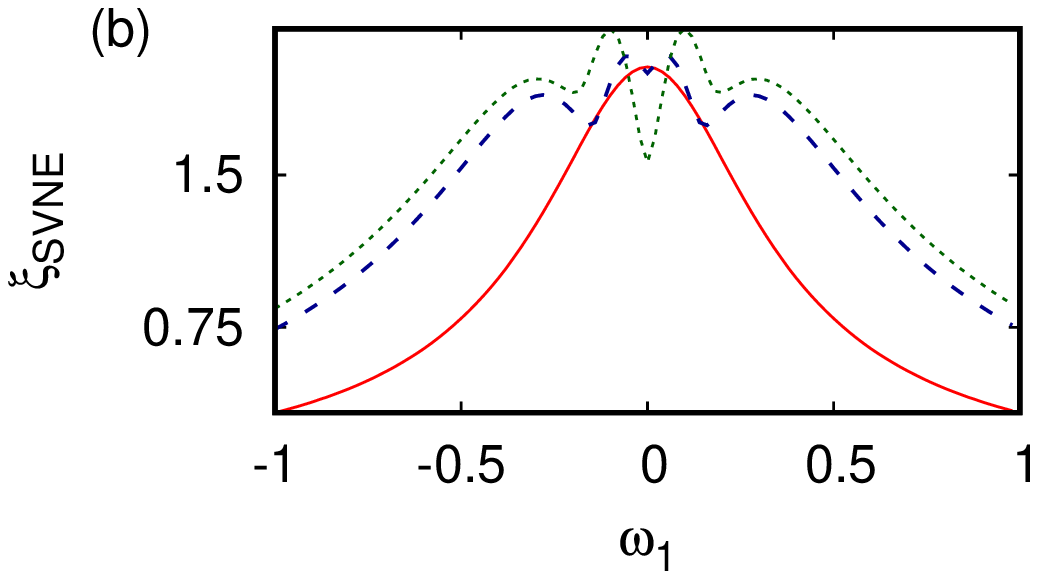}
\caption{(a) $E(N,k)$ vs.  $\omega_{1}$ for $N=4$ 
and $k=0,2,4$ in the BEC model. (b) $\xi_{\textsc{svne}}$ vs. $\omega_{1}$ for $N=4, \,k=0,1,2$.  The curves correspond to $k=0$ (red solid), $1$ (blue dashed), $2$ (green dotted) and $4$ (orange dot-dashed). $\lambda=0.25$.}
\label{fig:energySpectrumSVNE}
\end{figure}

In \fref{fig:energySpectrumSVNE}(a), 
 $E(N=4,k)$ is  plotted against  
$\omega_{1}$ for $k=0,2,4$, with 
$\lambda=0.25$. 
$E(N, N - k)$ is the reflection of  
$E(N, k)$  about the value 
$\omega_{0}N + UN^{2}$.  
Avoided  energy-level 
crossings are seen at $\omega_{1}=0$. In order to  
 set the reference level for the extent of entanglement between the two wells, we compute 
 $\xi_{\textsc{svne}} 
 = -\mathrm{Tr}\,
 (\rho_{\textsc{a}} \log_{2} \, \rho_{\textsc{a}})$, 
  where  $\rho_{\textsc{a}}$ 
is the reduced density matrix of the subsystem A. 
($\xi_{\textsc{svne}}$  is also equal to 
$-\mathrm{Tr}\,(\rho_{\textsc{b}} \log_{2} \, \rho_{\textsc{b}})$, since  
   $\ket{\psi_{N,k}}$ is a bipartite pure state.) 
Plots of $\xi_{\textsc{svne}}$ corresponding to the state $\ket{\psi_{4,k}}$ for $k = 0,1,2$ 
 are shown in 
 \fref{fig:energySpectrumSVNE}(b). 
 The states $\ket{\psi_{4,3}}$ and  
$\ket{\psi_{4,1}}$ have the same 
$\xi_{\textsc{svne}}$, (as do the states  
$\ket{\psi_{4,4}}$ and  $\ket{\psi_{4,0}}$), 
owing to the $k\leftrightarrow  N-k$ symmetry. 
It is evident that there is a significant extent of entanglement close to the avoided crossing,  and   $\omega_{1}=0$ is marked by a local maximum or minimum in $\xi_{\textsc{svne}}$. 

\begin{figure*}
\includegraphics[width=0.3\textwidth]{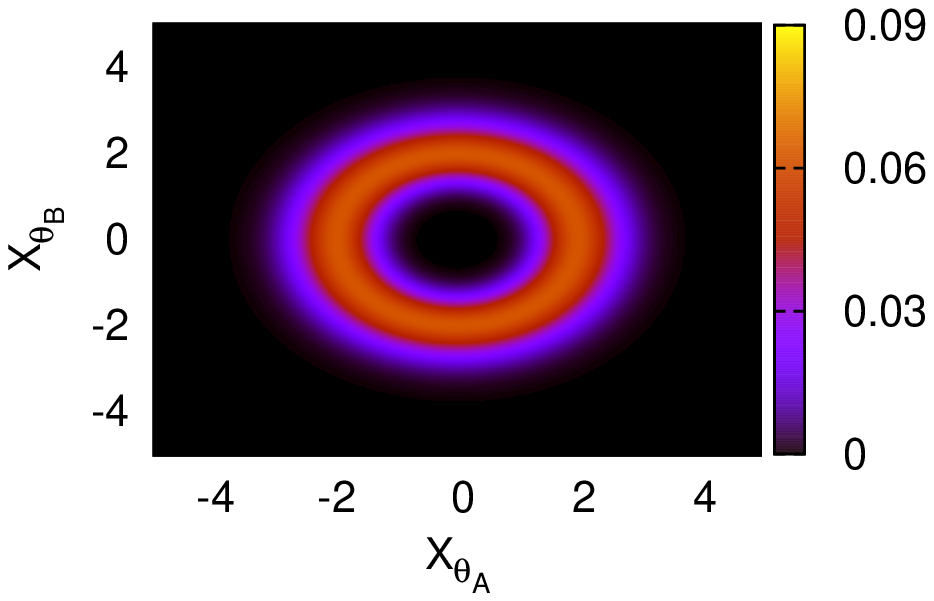}
\includegraphics[width=0.3\textwidth]{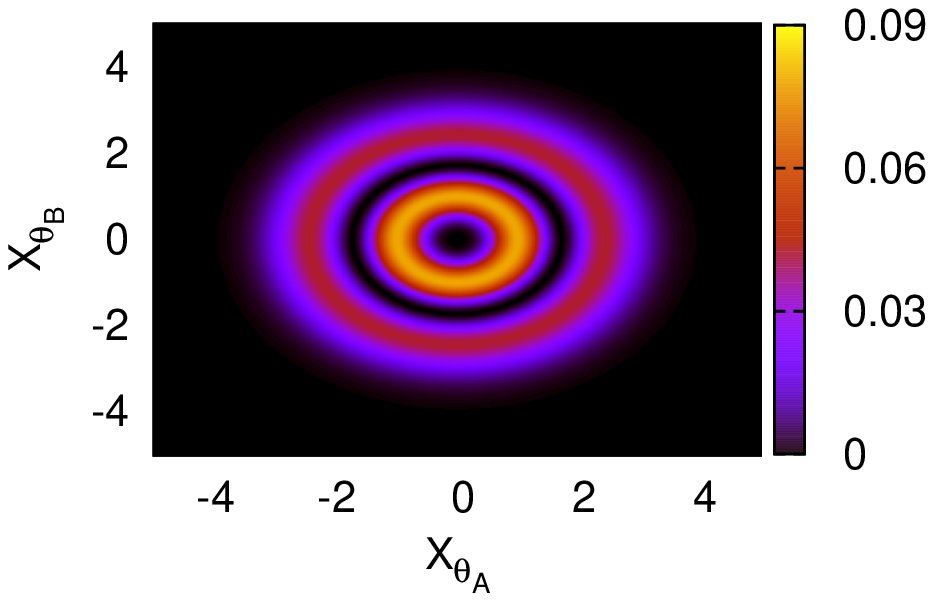}
\includegraphics[width=0.3\textwidth]{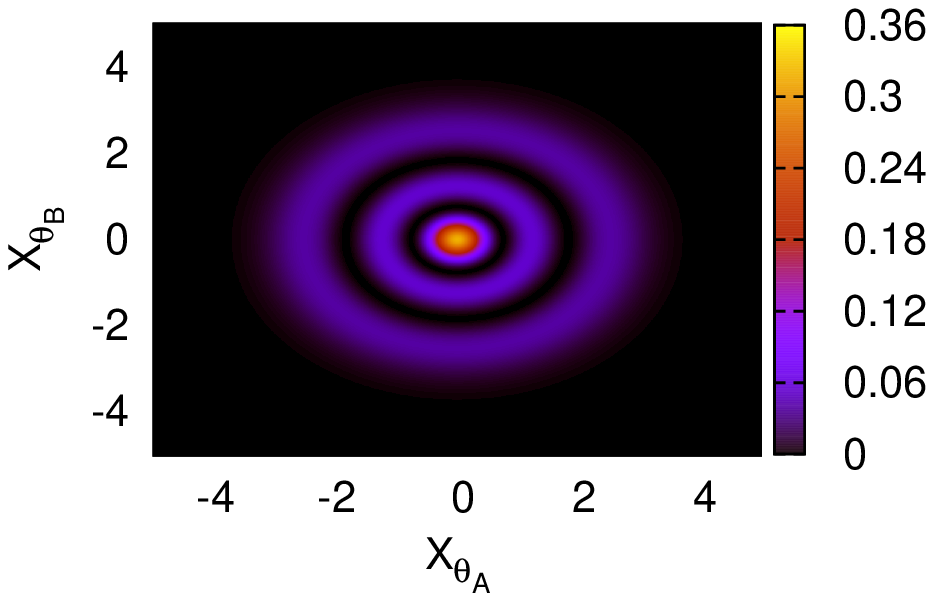}\\
\includegraphics[width=0.3\textwidth]{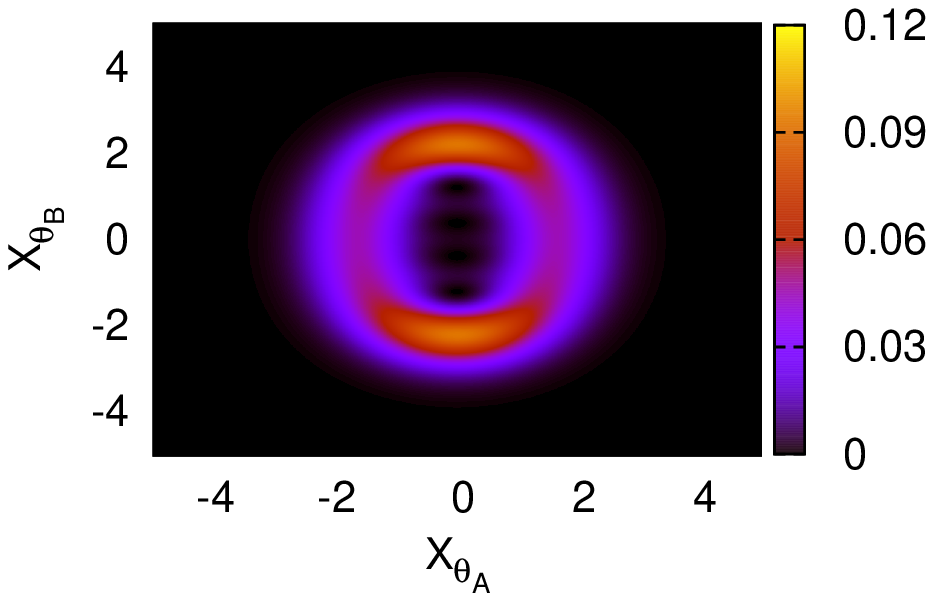}
\includegraphics[width=0.3\textwidth]{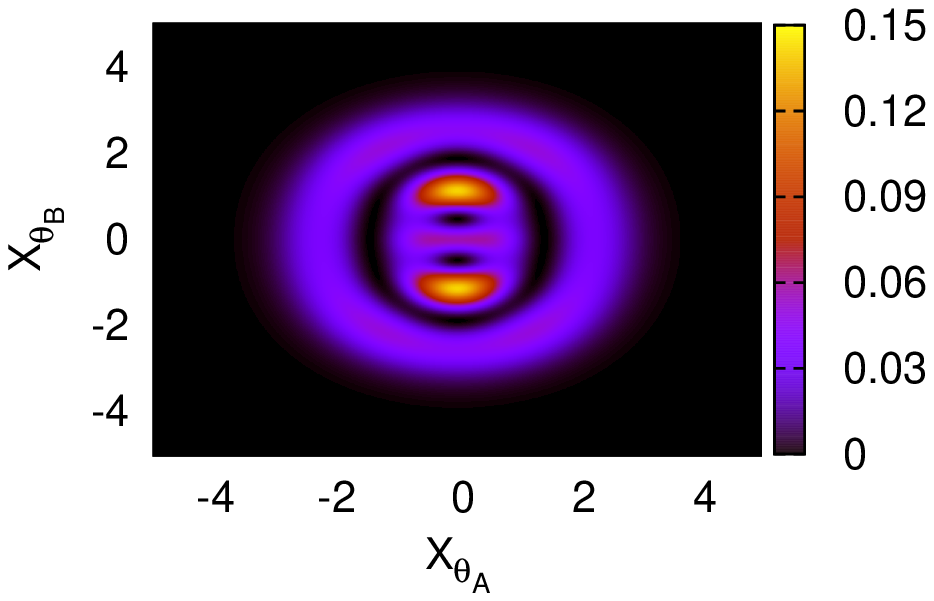}
\includegraphics[width=0.3\textwidth]{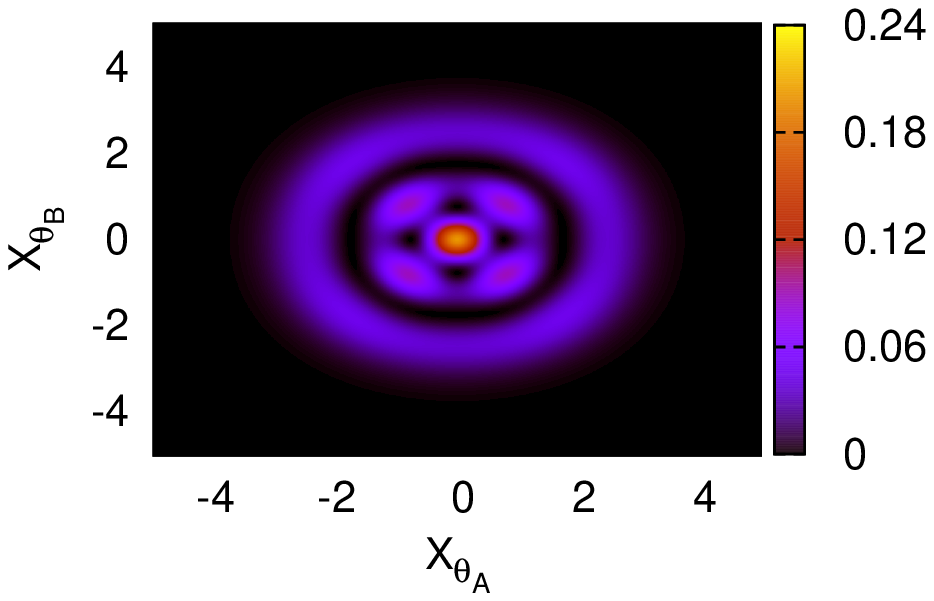}\\
\includegraphics[width=0.3\textwidth]{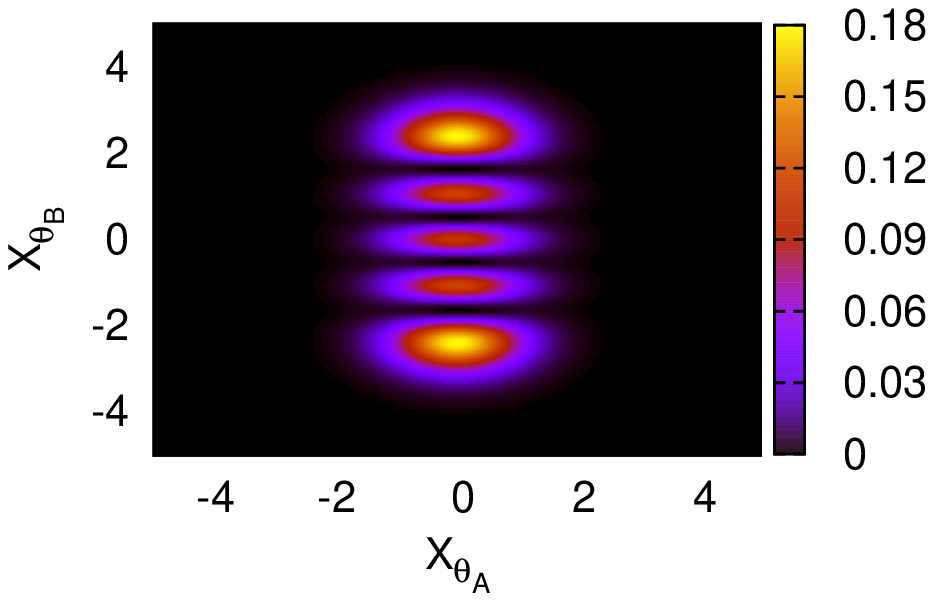}
\includegraphics[width=0.3\textwidth]{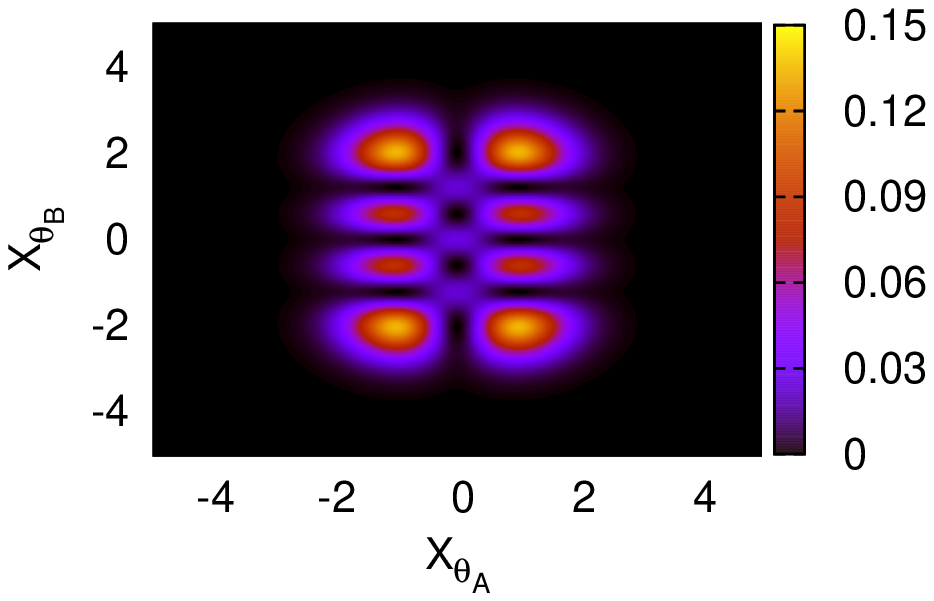}
\includegraphics[width=0.3\textwidth]{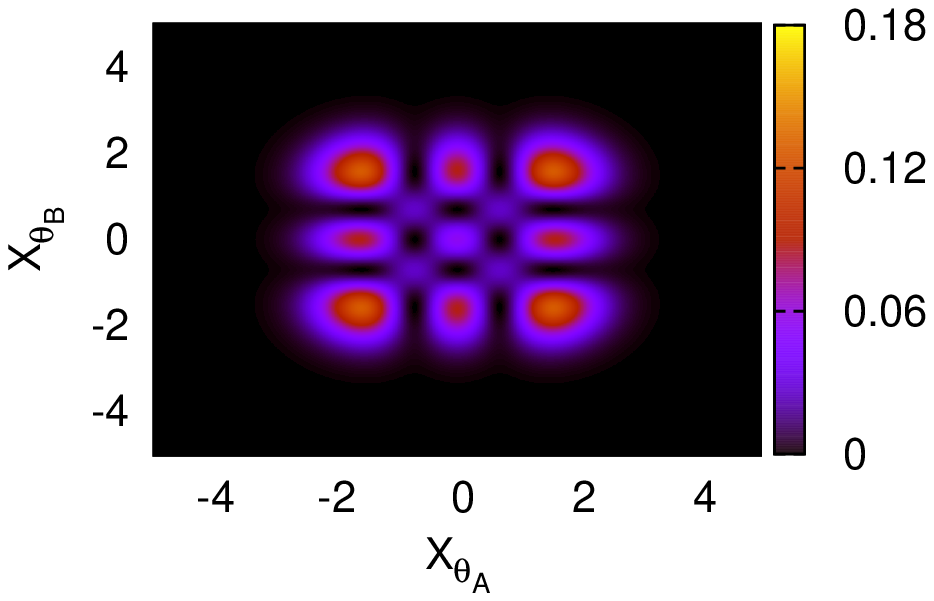}
\caption{$\thetaa=0,\thetab=\frac{1}{2}\pi$ slice of 
the tomogram for $N=4$ in the BEC model. Left to right,  $k=0, 1$ and 
$2$. Top to bottom,  $\omega_{1}=0, 0.1$ and $1$.}
\label{fig:tomosAC}
\end{figure*}

\Fref{fig:tomosAC} depicts   $\thetaa = 0, \thetab = \frac{1}{2}\pi$ sections of the tomograms corresponding to the states 
$\ket{\psi_{4,k}}$ for  $k=0,1, 2$ and   
 $\omega_{1}=0, 0.1, 1$.
It is clear that, for a given value of $\omega_{1}$, the qualitative features of the tomograms are altered considerably as $k$ is varied. 
The patterns in the tomograms also reveal   nonlinear correlations between the quadrature variables $X_{\thetaat}$ and $X_{\thetabt}$ (top panel). For  instance, the tomogram slice on the top right 
 shows a probability distribution that is essentially  unimodal and symmetric about the origin with the annular structures diminished in magnitude. It is 
clear that this case is less correlated than the tomogram  in the top left corner. This conforms to the observed trend in the extent of entanglement (compare $\xi_{\textsc{svne}}$ corresponding to $k=0$ and $k=2$ at $\omega_{1}=0$ in \fref{fig:energySpectrumSVNE} (b)). Again,  in the bottom panel of the figure, the sub-structures in the patterns increase with increasing $k$, signifying 
a higher degree of nonlinear correlation. This is in 
consonance  with the trend in the  entanglement at 
$\omega_{1}=1$ (\fref{fig:energySpectrumSVNE} (b)). We therefore expect $\epsarg{tei}$ 
and its 
averaged version   
$\xi_{\textsc{tei}}$  to be 
 much better entanglement indicators  than 
  $\epsarg{pcc}$ and $\xi_{\textsc{pcc}}$. 
We also  mention here that the current experimental techniques of testing CV entanglement based on the variances and covariances of suitably chosen observables~\cite{ExptCVentang} are  
not as effective as calculating nonlinear 
correlators,  for the same reason.

\begin{figure}
\includegraphics[width=0.5\textwidth]{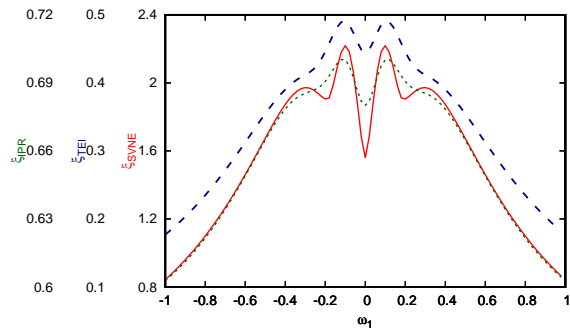}
\caption{$\xi_{\textsc{svne}}$ (red solid line), $\xi_{\textsc{tei}}$ (blue dashed line) and $\xi_{\textsc{ipr}}$ (green dotted line) vs. 
$\omega_{1}$, for the state 
$\ket{\psi_{4,2}}$ in the BEC model.}
\label{fig:N4k2VaryOmega1}
\end{figure}

Our detailed investigations reveal that 
$\xi_{\textsc{tei}}$ and $\xi_{\textsc{ipr}}$ 
follow the trends in 
$\xi_{\textsc{svne}}$ 
reasonably well for generic eigenstates of $H_{\textsc{bec}}$.
This is illustrated in 
\fref{fig:N4k2VaryOmega1}, which shows plots of these indicators as functions of $\omega_{1}$. 
Apart from examining the suitability of  
$\epsarg{pcc}$ as an  entanglement indicator, 
 we have also checked for the extent of linear correlation between any two indicators based on 
 the corresponding PCC, as follows. 
  We have obtained  $100$ values each of $\xi_{\textsc{tei}}$ and $\xi_{\textsc{svne}}$ for different values of $\omega_{1}$ in the range 
  $(-1, 1)$ 
  in steps of $0.02$. Treating the two sets of values    
  as two sets of random numbers, we obtain the PCC between them, as defined 
   in Eq. \eref{eqn:PCC}. The  PCC between $\xi_{\textsc{tei}}$ and $\xi_{\textsc{svne}}$ (respectively,  $\xi_{\textsc{ipr}}$ and $\xi_{\textsc{svne}}$) estimates the extent of  linear correlation between the two indicators, and 
is found to be $0.97$ (resp., $0.99$) in the case shown in \fref{fig:N4k2VaryOmega1} corresponding to $\ket{\psi_{4,2}}$.  
 (In general, the  PCC ranges from $1$ for complete correlation,  to $-1$ for  maximal anti-correlation. 
 Its vanishing indicates the absence of  linear correlation).  

\begin{figure}
\includegraphics[width=0.3\textwidth]{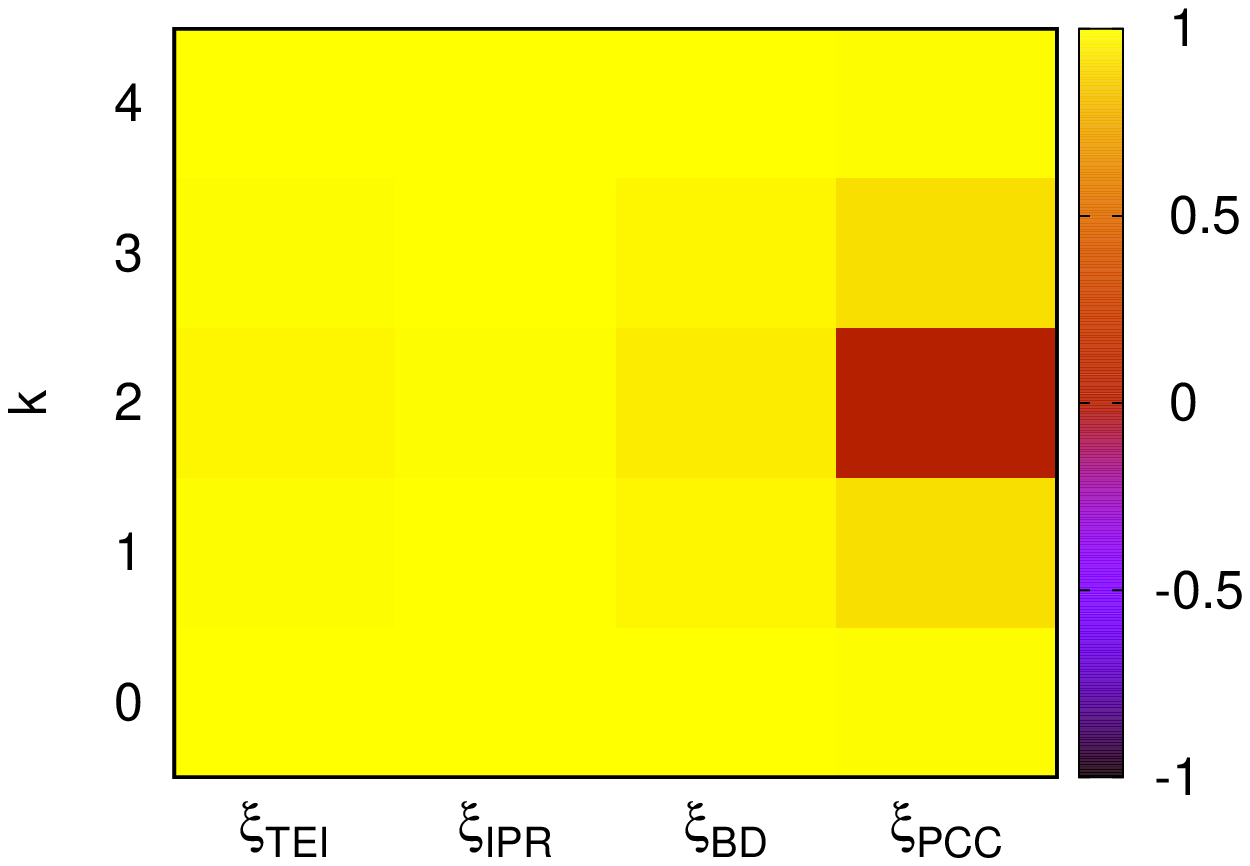}
\includegraphics[width=0.3\textwidth]{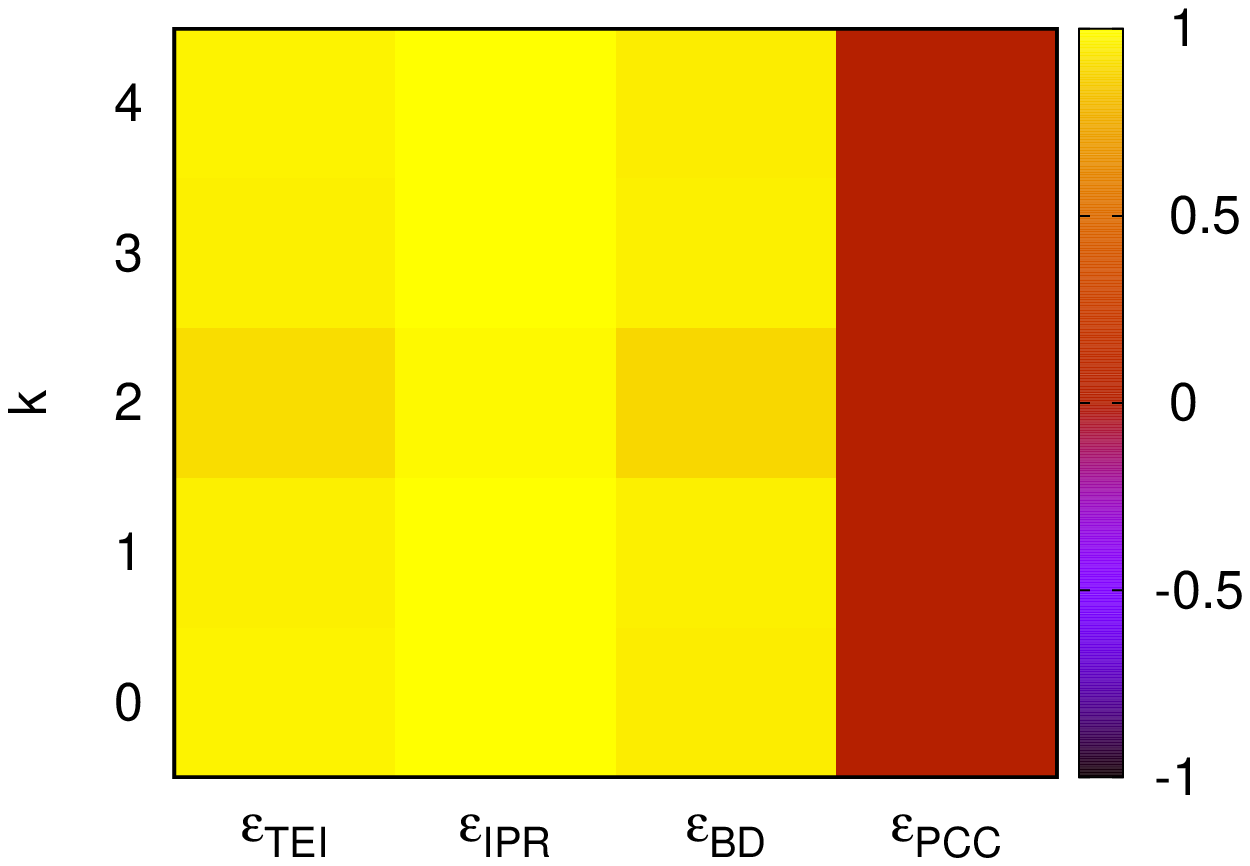}
\includegraphics[width=0.3\textwidth]{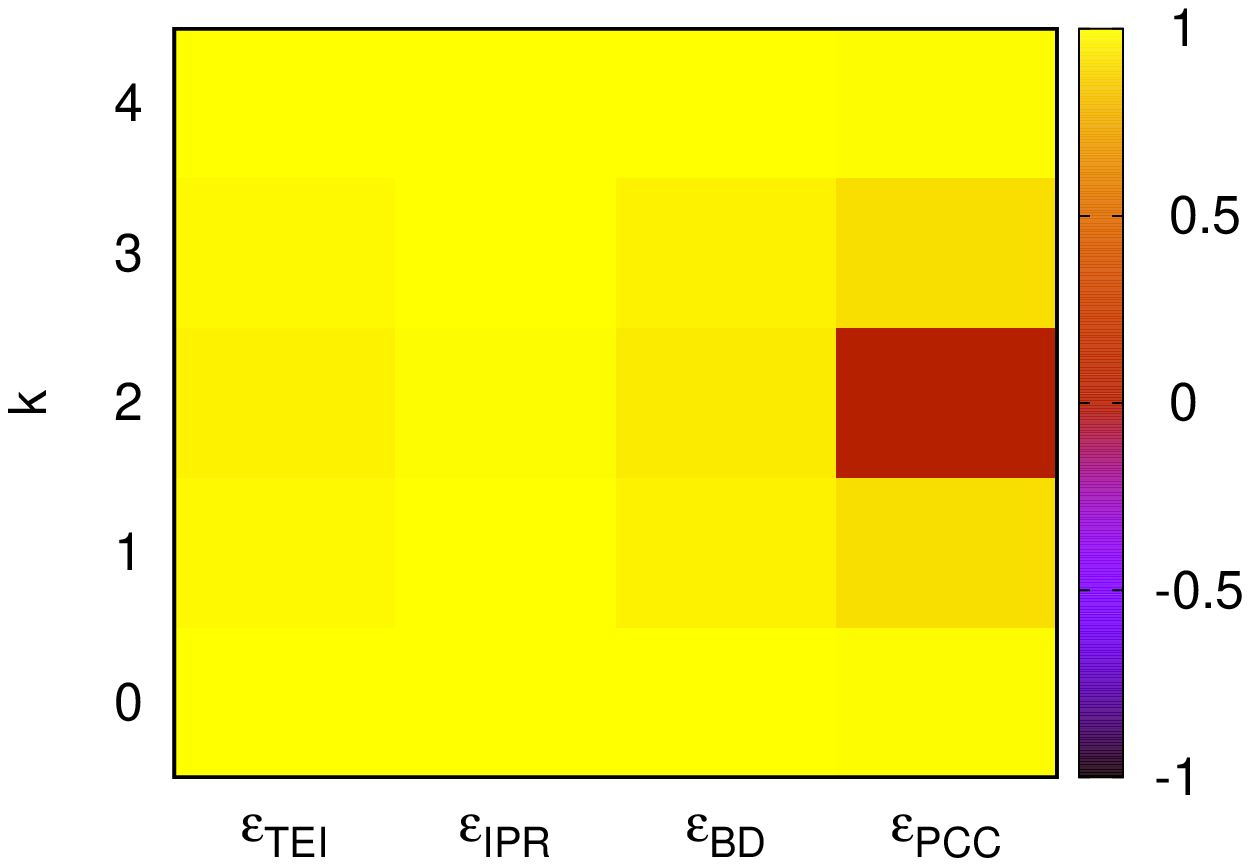}
\caption{Correlation of $\xi_{\textsc{svne}}$ 
with 
 $\xi$-indicators (left),   
 with $\neweps$-indicators for the slice 
 $\thetaa=0,\thetab=\frac{1}{2}\pi$ (centre), 
 and with 
 $\neweps$-indicators  for the slice 
 $\thetaa=0,\thetab=0$ (right), for  the 
 eigenstates $\ket{\psi_{4,k}}, 
 \,0 \leqslant k \leqslant 4$ in the BEC model.}
\label{fig:CorrPlotN4VaryOmega1}
\end{figure}

Figure \ref{fig:CorrPlotN4VaryOmega1} 
shows 
the PCC  between 
$\xi_{\textsc{svne}}$ and various indicators, 
for the eigenstates 
$\ket{\psi_{4,k}}$ where $k=0,1,2,3,4$.
From \fref{fig:CorrPlotN4VaryOmega1}(a), we see that $\xi_{\textsc{ipr}}$,  $\xi_{\textsc{tei}}$ 
and $\xi_{\textsc{bd}}$ are very good  entanglement indicators. We have also found that 
all these indicators improve with increasing  $N$.
The performance of the $\neweps$-indicators depends, of course,  on the specific choice of the tomographic section. For instance, 
$\epsarg{tei}$ and $\epsarg{bd}$ perform marginally better for the slice  
$\thetaa = 0, \,\thetab = 0$ 
than for the slice
$\thetaa = 0, \,\thetab = \frac{1}{2}\pi$. 
It is also evident that 
$\xi_{\textsc{pcc}}$ does not fare as well as the other indicators. This is to be  expected, since 
 $\xi_{\textsc{pcc}}$  only captures  linear correlations, as already emphasised.

   We have verified that the sensitivity  of all the indicators decreases with an increase in 
   $\lambda$, the strength of  
 the coupling between the two  subsystems   
 (as in Eq. \ref{eqn:HBEC}).   
$\xi_{\textsc{ipr}}$, however, 
 remains closer to 
 $\xi_{\textsc{svne}}$ 
than the other indicators. 
This fact is consistent with inferences drawn from our earlier work~\cite{sharmila2}  
about  the relation between 
the Hamming distance~\cite{HammingInQM}  
and the efficacy of $\xi_{\textsc{ipr}}$. 
We recall that the 
Hamming distance between two bipartite qudits  
$\ket{u_{1}}\otimes \ket{u_{2}}$ and $\ket{v_{1}}\otimes\ket{v_{2}}$ attains its maximum value of 
$2$ when $\aver{u_{1}|v_{1}}=0$ and $\aver{u_{2}|v_{2}}=0$. 
A straightforward extension to CV systems implies  that the Hamming distance between $\ket{k_{1},N-k_{1}}$ and $\ket{k_{2},N-k_{2}}$ is 2 (so that these states are Hamming-uncorrelated), if $k_{1}\neq k_{2}$. Participation ratios are valid measures of entanglement for superpositions of Hamming-uncorrelated states in spin systems~\cite{ViolaBrown}. We have demonstrated in our earlier work  that $\xi_{\textsc{ipr}}$ effectively mimics standard measures of entanglement in  CV systems as well. In the present instance, the eigenstates $\ket{\psi_{N,k}}$ are superpositions of the states $\lbrace \ket{j,N-j} \rbrace$ which are Hamming-uncorrelated for different values of 
$j$. This is the reason for the usefulness   
of $\xi_{\textsc{ipr}}$ as an entanglement indicator even for larger values of $\lambda$. 

\begin{figure}
\includegraphics[width=0.45\textwidth]{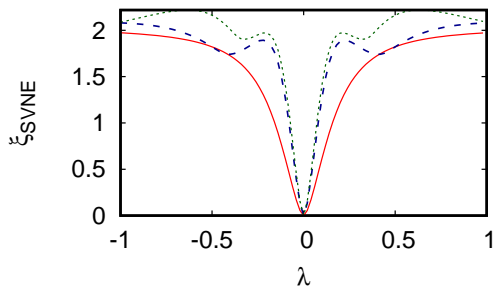}
\caption{$\xi_{\textsc{svne}}$ vs. $\lambda$ for $N=4, \,k=0,1,2$, for the BEC model.  The curves correspond to $k=0$ (red solid), $1$ (blue dashed) and $2$ (green dotted). $\omega_{1}=0.25$. }
\label{fig:energySpectrumSVNEvaryLambda}
\end{figure}

We now proceed to examine 
quantitatively the efficacy of the entanglement indicators  as functions of  $\lambda$. For numerical computation we have set $\omega_{1}=0.25$.  
Consider, as an illustration, plots of the eigenvalues 
$E(4,k)$ ($k=0,2,4$) as functions of  $\lambda$.  
These plots are  exactly the 
same as those in 
\fref{fig:energySpectrumSVNE}(a), with 
$\omega_{1}$ replaced by $\lambda$ on the 
horizontal axis, since 
$E(N, k)$ only depends on the parameters 
$\omega_{1}$ and $\lambda$ in the 
symmetric combination $\lambda_{1} = (\lambda^{2} 
+ \omega_{1}^{2})^{1/2}$. 
The avoided crossing of energy levels now occurs at 
$\lambda = 0$.   
But this symmetry between 
$\omega_{1}$ and 
$\lambda$ does not extend to 
the unitary transformation $V$, and hence 
to  the eigenstates 
of $H_{\textsc{bec}}$. 
(Recall that $V$ involves the parameter 
$\kappa = \tan^{-1}(\lambda/\omega_{1})$.) 
When  $\lambda=0$ 
there is  no linear interaction between the two modes.  $V$ then  reduces to the identity operator, and 
$H_{\textsc{bec}}$ is 
 diagonal in the basis 
$\lbrace \ket{k,N-k} \rbrace$. 
We therefore  expect the entanglement to vanish  at the avoided crossing. This is borne out in \fref{fig:energySpectrumSVNEvaryLambda} 
 in which  
$\xi_{\textsc{svne}}$   
for the state $\ket{\psi(4,k)}$ is plotted  
for different values of $k$. 
As before,  it suffices to depict the cases  
$k = 0,1$ and $2$ because of the $k \leftrightarrow N-k$ 
symmetry. We observe that, in the 
case $k = 0$, 
 while there is a  minimum 
in $\xi_{\textsc{svne}}$ at $\lambda = 0$, 
there is a maximum in this quantity at 
$\omega_{1} = 0$  
(\fref{fig:energySpectrumSVNE}(b)).  

We have also calculated the 
PCC between various indicators and $\xi_{\textsc{svne}}$ for the set of states $\ket{\psi_{4,k}}, \,0\leqslant k \leqslant 4$, 
using $100$ values of each of 
the $\xi$-indicators 
calculated for each 
$\lambda$ in the range $[-1, 1]$ with a step size of 
$0.02$. The results are   very similar to those 
already found 
(see figure \ref{fig:CorrPlotN4VaryOmega1})
using  $\omega_{1}$ as the variable 
parameter instead of $\lambda$.

\subsection{Atom-field interaction model}

We turn next to the case of a multi-level atom (modelled by an anharmonic oscillator) that is  linearly coupled with strength $g$ to a radiation field 
of  frequency $\omega_{\mathrm{f}}$. The effective Hamiltonian (setting $\hbar = 1$ )  is given by~\cite{agarwalpuri}
\begin{equation}
H_{\textsc{af}}=\omega_{\mathrm{f}} a^{\dagger} a +
 \omega_{\mathrm{a}} b^{\dagger} b + \gamma b^{\dagger \,2} b^{2} + g ( a^{\dagger} b + a b^{\dagger}).
\label{eqn:HAF}
\end{equation}
 $\omega_{\mathrm{a}}$ and 
 $\gamma$ ($> 0$ for stability) are 
 constants. 
  $(a,a^{\dagger})$ 
 and $(b,b^{\dagger})$ are the annihilation and creation operators for the field mode  and the oscillator mode, 
 respectively. As before, 
  $N_{\mathrm{tot}} = a^{\dagger}a+b^{\dagger}b$ and $[H_{\textsc{af}},N_{\mathrm{tot}}]=0$.  
  As in the BEC model 
  of the preceding section, the eigenvalues $E_{\af}(N,k)$ and the common eigenstates $\ket{\phi_{N,k}}$ of these two operators are labelled by  $N = 0,1,\ldots$ (the eigenvalue of 
$N_{\mathrm{tot}}$) and, within each $(N+1)$-dimensional 
subspace for a given $N$, by the index $k$ that runs from $0$ to $N$. 

\begin{figure}
\includegraphics[width=0.45\textwidth]{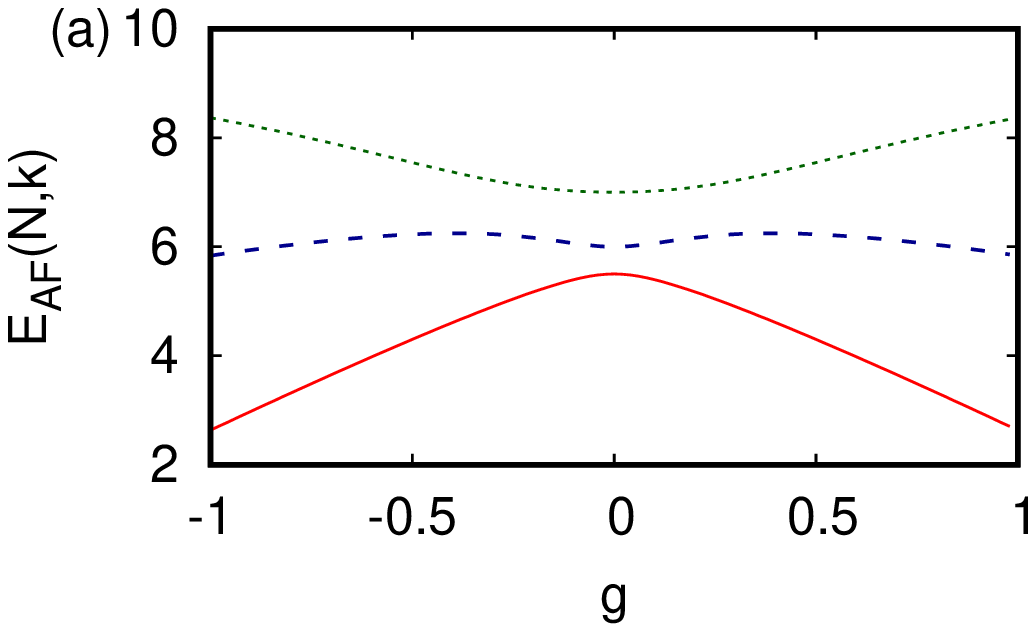}
\includegraphics[width=0.45\textwidth]{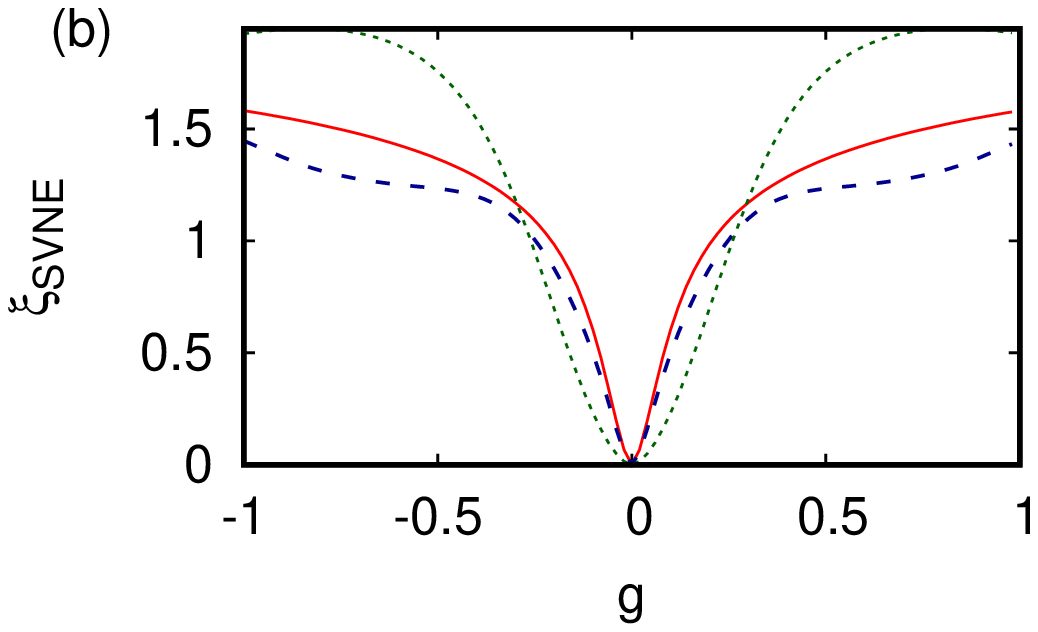}
\caption{(a) $E_{\af}(N,k)$ and (b) $\xi_{\textsc{svne}}$ vs. $g$ for $N=4$, $k=0,1,2$ in the atom-field interaction model. The curves correspond to $k=0$ (red solid), $1$ (blue dashed) and $2$ (green dotted). $\omega_{\mathrm{f}}=1.5, \,\omega_{\mathrm{a}}=1, \,
\gamma=1$.}
\label{fig:energySpectrumSVNEagarwalVaryLambdaDetuned}
\end{figure}
 We find 
 $\ket{\phi_{N,k}}$ and 
 $E_{\af}(N,k)$ numerically. 
Figures \ref{fig:energySpectrumSVNEagarwalVaryLambdaDetuned}(a) and (b) show plots of  
$E_{\af}(N,k)$ and $\xi_{\textsc{svne}}$ versus $g$  for $N=4$  and  $k=0,1,2$ in the case   
$\omega_{\mathrm{f}}=1.5, \,\omega_{\mathrm{a}}=1$. 
Avoided crossings occur at $g=0$, 
with a corresponding  minimum in  
$\xi_{\textsc{svne}}$ 
that drops down to zero 
for each of the three states 
$\ket{\phi_{4,0}},  \ket{\phi_{4,1}}$ and 
$\ket{\phi_{4, 2}}$. These states are 
therefore unentangled at $g=0$, i.e., 
in the absence of interaction between the 
two modes of the bipartite system, as one might expect.  

In order to examine what happens when there 
{\em  is} a 
crossing of energy levels, we introduce a degeneracy 
by setting  $\omega_{\mathrm{f}}= \omega_{\mathrm{a}}$.
\begin{figure}
\includegraphics[width=0.45\textwidth]{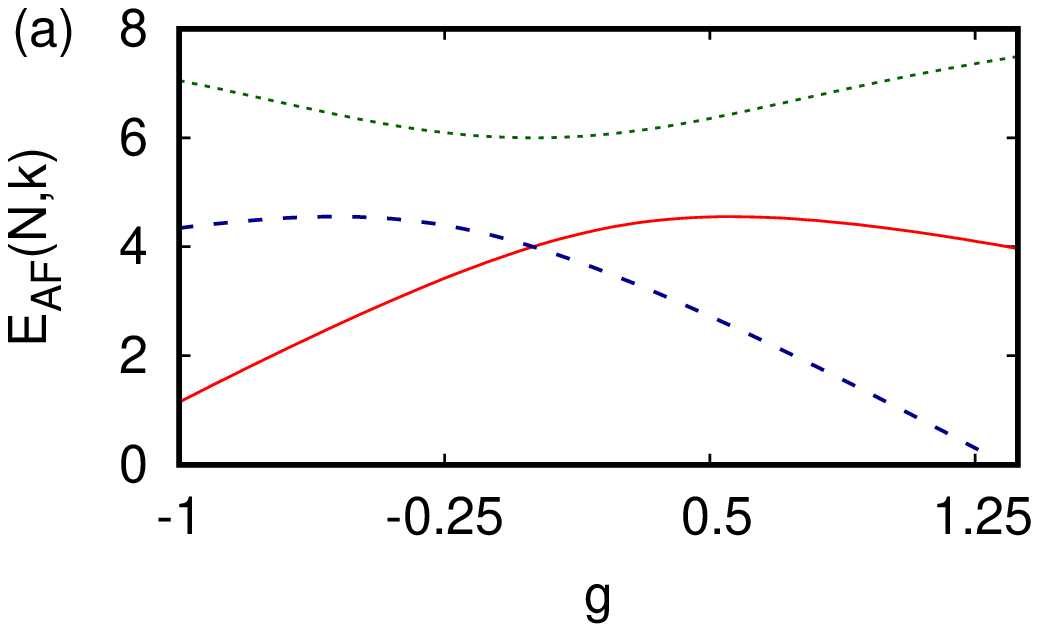}
\includegraphics[width=0.45\textwidth]{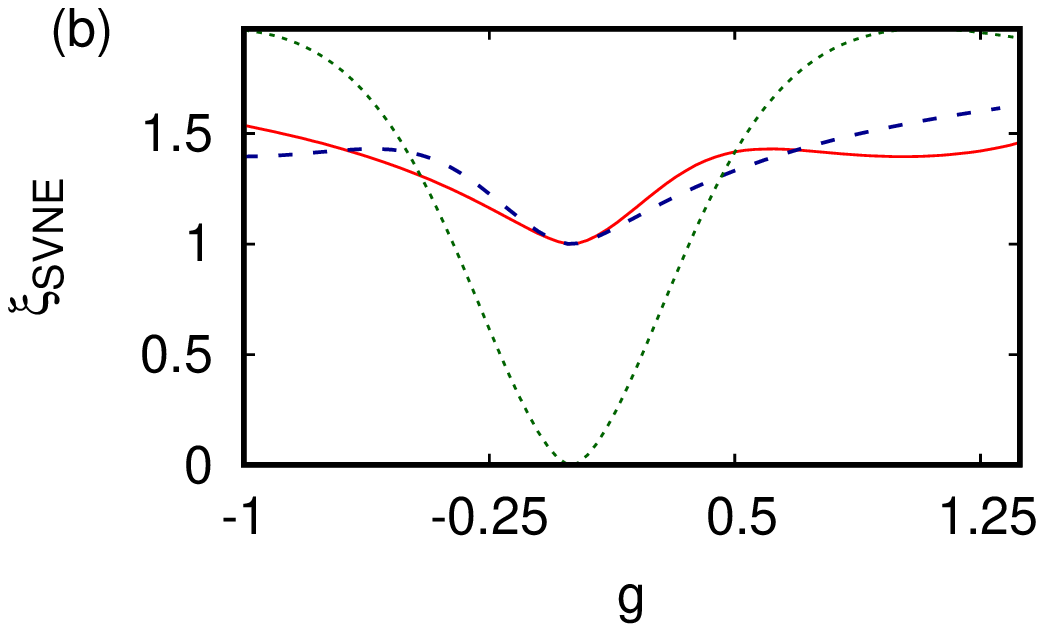}
\caption{(a) $E_{\af}(N,k)$ and (b) $\xi_{\textsc{svne}}$ vs. $g$ for $N=4$, $k=0,1,2$ in the atom-field interaction model, in the degenerate case  
$\omega_{\mathrm{f}}=\omega_{\mathrm{a}}
= 1$. The curves correspond to $k=0$ (red solid), $1$ (blue dashed) and $2$ (green dotted). 
$\gamma=1$.}
\label{fig:energySpectrumSVNEagarwalVaryLambda}
\end{figure}
Figures \ref{fig:energySpectrumSVNEagarwalVaryLambda} (a) and (b) are plots of  $E_{\af}(N,k)$ and  $\xi_{\textsc{svne}}$  versus $g$ for $N=4$ and 
$k = 0,1, 2$, with  
$\gamma$, $\omega_{\mathrm{a}}$ and $\omega_{\mathrm{f}}$  set equal to $1$.
Both a level crossing and an avoided crossing are seen to occur at $g=0$, signalled by a 
minimum in $\xi_{\textsc{svne}}$ for 
each of the three states concerned.  The crossing 
of $E_{\af}(4,0)$ and $E_{\af}(4,1)$ arises as follows. 
Let  $\ket{p,4-p}$ denote the product state  $\ket{p}\otimes\ket{4-p}$, where  $\ket{p}$ is a photon number state of the field mode and $\ket{4-p}$ is an oscillator 
state of the atom mode. 
When $\gamma = \omega_{\mathrm{a}} 
= \omega_{\mathrm{f}} = 1$ and $g = 0$, the Hamiltonian 
reduces to $a^{\dagger} a + (b^{\dagger} b)^{2}$. 
The energy levels
$E_{\af}(4,0)$ and $E_{\af}(4,1)$ 
become degenerate 
at the value $4$. 
The degeneracy occurs because the 
operator $\ket{4,0}\bra{3,1} + \ket{3,1}\bra{4,0}$ 
commutes with 
$H_{\af}$ when  $\omega_{\mathrm{a}}=\omega_{\mathrm{f}}$ and $g=0$.
Mixing of the states  
$\ket{4,0}$ and  $\ket{3,1}$
occurs, and the corresponding energy eigenstates are given by 
the symmetric linear combination 
$\ket{\phi_{4,0}} =
(\ket{4,0}+\ket{3,1})/\sqrt{2}$ 
and the antisymmetric linear combination 
$\ket{\phi_{4,1}} =  (\ket{4,0}-\ket{3,1})/\sqrt{2}$.  
As the symmetries of the two states are different, 
 the level crossing does not violate the 
von Neumann-Wigner no-crossing theorem. 
At the crossing,  each of the  states 
$\ket{\phi_{4,0}}$ and 
$\ket{\phi_{4,1}}$ 
remains a manifestly  
 entangled state that is,  in fact, a Bell state.  
This is why the corresponding 
$\xi_{\textsc{svne}}$ does not vanish at that point, but merely dips to a local minimum with value $1$, 
characteristic of a Bell state.
It is interesting to note that  the degeneracy  
that occurs when 
$\omega_{\mathrm{f}}=\omega_{\mathrm{a}}$ 
ensures entanglement even in the absence of 
any interaction between the two modes.  

The level
 $E_{\af}(4,2)$, on the other hand,  is repelled and has the value 
 $6$ at $g =  0$. The corresponding eigenstate 
 $\ket{\phi_{4,2}}$ becomes 
 the unentangled product state $\ket{2,2}$ at the 
 avoided crossing, and 
 $\xi_{\textsc{svne}}$ drops to zero in this case, as expected.

\begin{figure}
\includegraphics[width=0.3\textwidth]{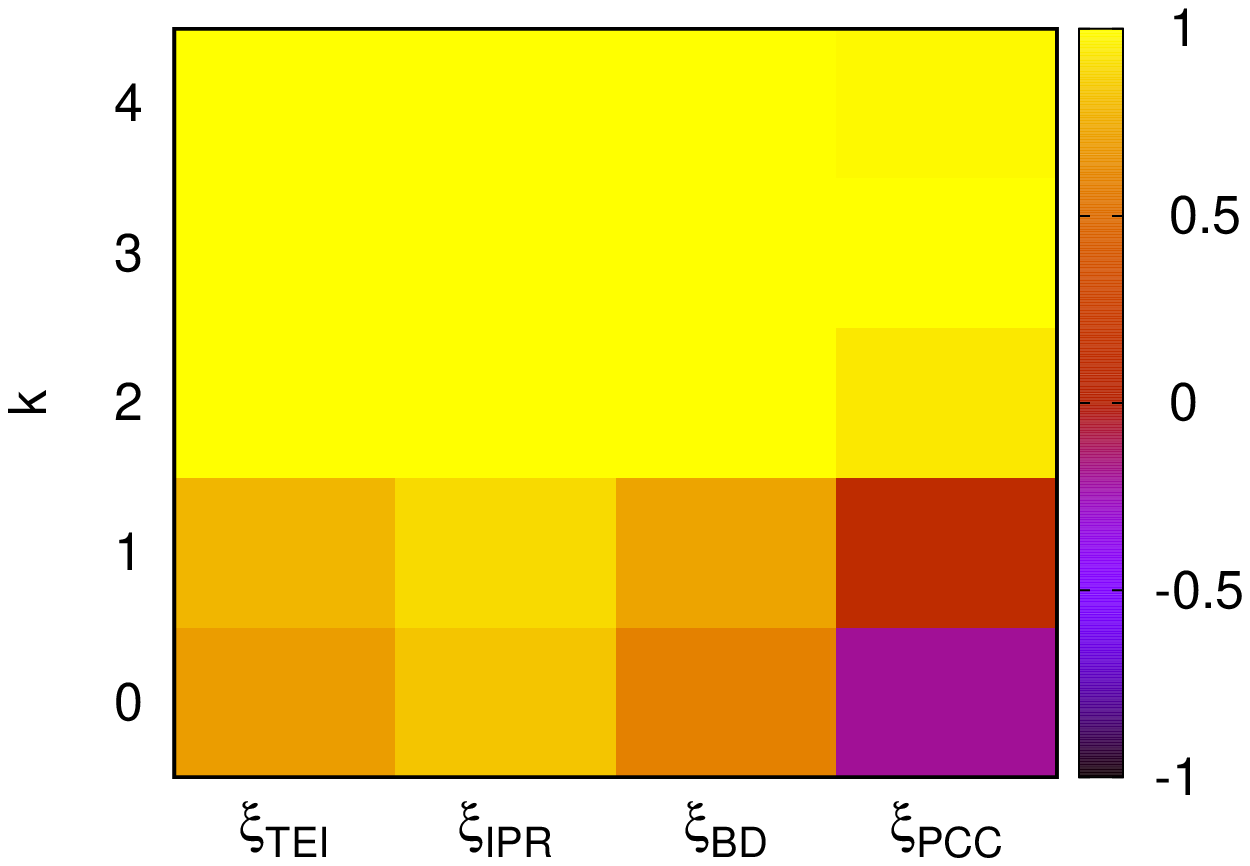}
\includegraphics[width=0.3\textwidth]{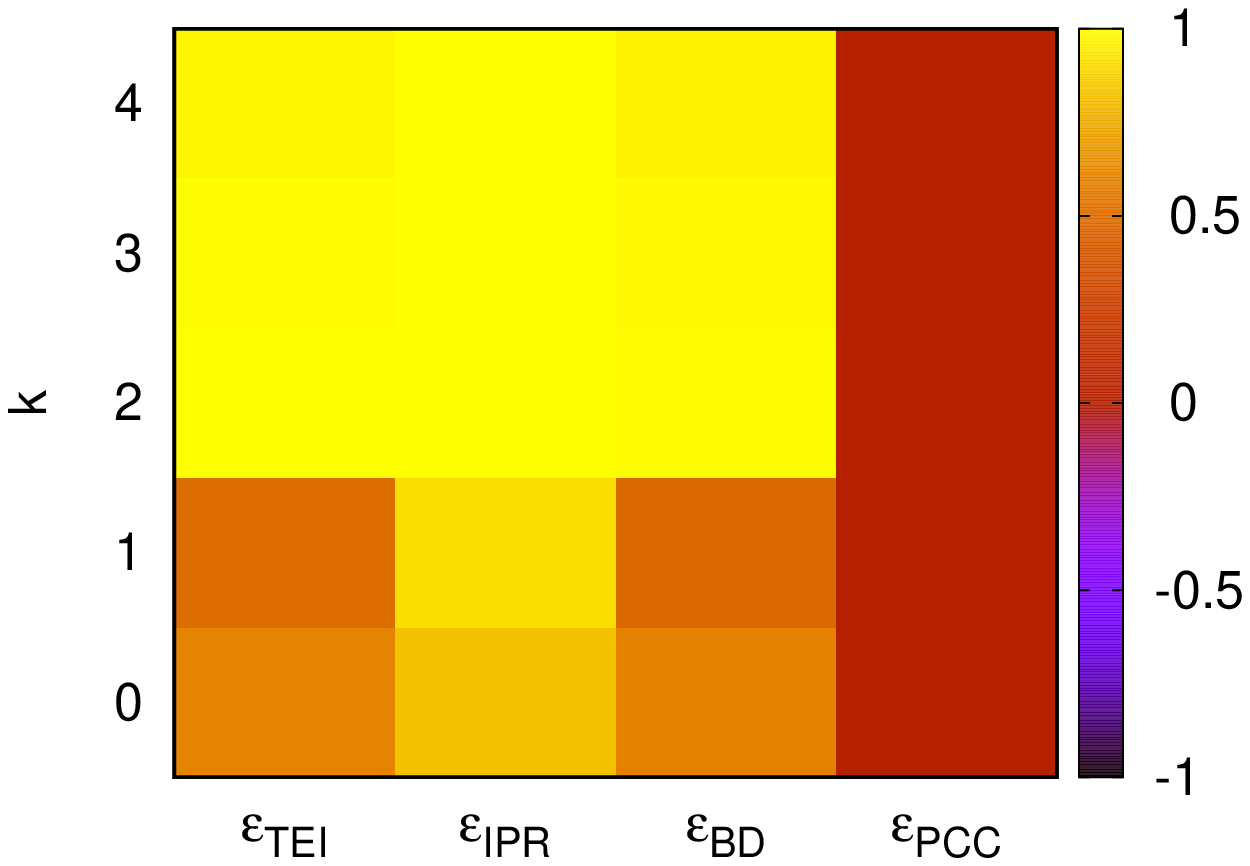}
\includegraphics[width=0.3\textwidth]{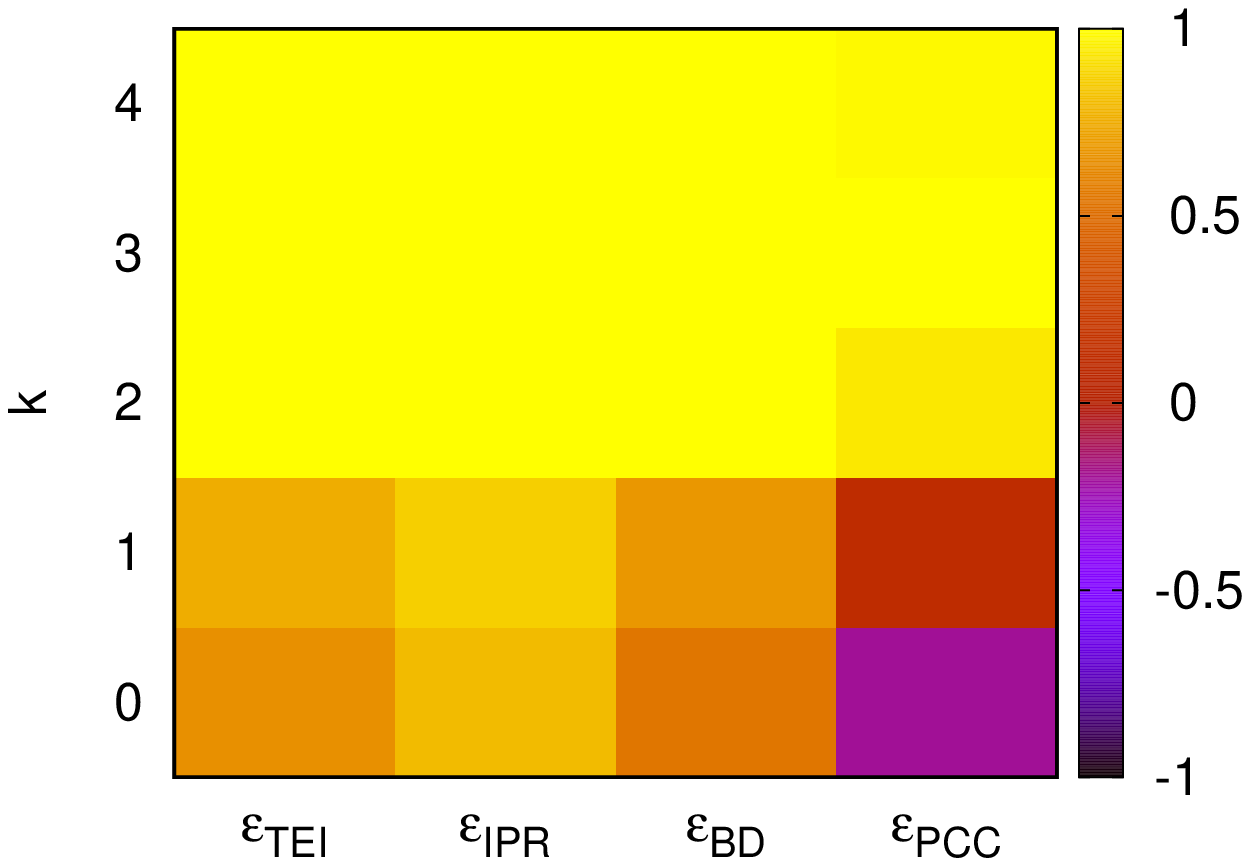}
\caption{Correlation of $\xi_{\textsc{svne}}$ 
with 
 $\xi$-indicators (left),   
 with $\neweps$-indicators for the slice 
 $\thetaa=0,\thetab=\frac{1}{2}\pi$ (centre), 
 and with 
 $\neweps$-indicators  for the slice 
 $\thetaa=0,\thetab=0$ (right), for the eigenstates $\ket{\phi_{4,k}}, 
 \,0 \leqslant k \leqslant 4$  in the atom-field interaction model.
 $\omega_{\mathrm{f}}=\omega_{\mathrm{a}}=
 \gamma=1$.}
\label{fig:CorrPlotN4agarwalVaryLambda}
\end{figure}

In \fref{fig:CorrPlotN4agarwalVaryLambda}, we plot 
the correlation between various indicators and 
$\xi_{\textsc{svne}}$. 
For  this purpose,  $80$ values of each of the $\xi$-indicators were calculated  
with  $g$ varied in the range $[-1,1.4]$ in  steps of 
$0.03$. Treating these as sets of random numbers, we obtain the PCC between the various indicators and 
$\xi_{\textsc{svne}}$,  as described in the foregoing. 
The performance of the entanglement indicators in this case is similar to that found  in the BEC system. Increasing $\gamma$ marginally decreases the efficacy of all the indicators.

\section{\label{sec:work2}Avoided crossings in multipartite HQ systems}
\subsection{Tavis-Cummings model}

As our third and final example,  we consider hybrid quantum systems comprising several qubits interacting with an external field.  
These systems are described by the class of  Tavis-Cummings models~\cite{tavis} in  a variety of  diverse physical situations  which include inherent field nonlinearities and inter-qubit interactions. The model we consider below is generic, applicable to a system of  several two-level atoms with nearest-neighbour couplings interacting with an external radiation field in the presence of  a   Kerr-like nonlinearity, or to a  chain of $M$ superconducting qubits interacting with a 
microwave field of frequency $\Omega_{\mathrm{f}}$. 
In the latter case, the model Hamiltonian (setting $\hbar = 1$) is given  by~\cite{QuantMeta2,QuantMeta} 
\begin{eqnarray}
\nonumber H_{\textsc{tc}}=\Omega_{\mathrm{f}} 
a^{\dagger} a + \chi a^{\dagger \,2} a^{2}  + \sum_{p=1}^{M} &\Omega_{p} \sigma_{p z} + \Lambda ( a^{\dagger} \sigma_{p}^{-} + a \sigma_{p}^{+})\\
&+ \sum_{p=1}^{M-1} \Lambda_{s} ( \sigma_{p}^{-} \sigma_{(p+1)}^{+} + \sigma_{(p+1)}^{-} \sigma_{p}^{+}).
\label{eqn:HTC}
\end{eqnarray}
Here, $\chi$ is the strength of the field nonlinearity, 
$\Lambda$ is the coupling strength between the 
field and each of the $M$ qubits, 
$\sigma_{p}^{\pm}$ are the ladder operators of the 
$p^{\rm th}$ qubit, and $\Lambda_{s}$  is the strength of the interaction between nearest-neighbour qubits. 
$\Omega_{p} = (\Delta_{p}^{2} + \epsilon^{2})^{1/2}$ 
 is the energy difference   between the two levels of the $p^{\rm th}$ qubit, where   $\Delta_{p}$ is the inherent excitation gap and $\epsilon$  is the detuning of the external magnetic flux from the flux quantum 
 $h/(2 e)$. In our numerical computations we have 
 used the  experimentally  relevant~\cite{QuantMeta}  
 parameter values $\Omega_{\mathrm{f}}/(2 \pi)
 = 7.78 \,{\rm   GHz}$ and  $\epsilon/(2 \pi) =  4.62\,
 {\rm  GHz}$. 
 The  level separations  $\Delta_{p}$  of the individual qubits have been  
 drawn from a  Gaussian  distribution with a mean given by 
  $\aver{\Delta}/(2 \pi) =  5.6\,{\rm  GHz}$ and a standard 
  deviation  $0.2 \,\aver{\Delta}$.

\begin{figure}
\includegraphics[width=0.4\textwidth]{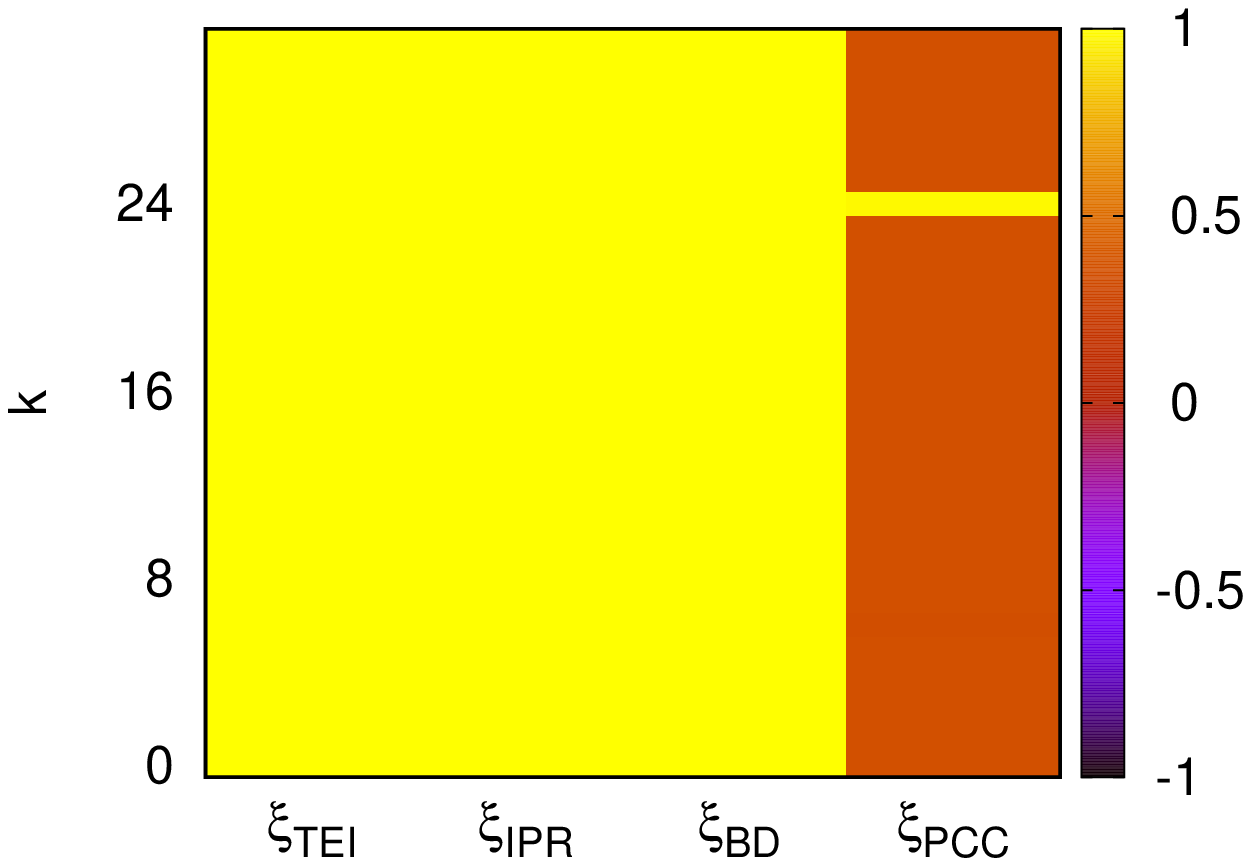}
\includegraphics[width=0.4\textwidth]{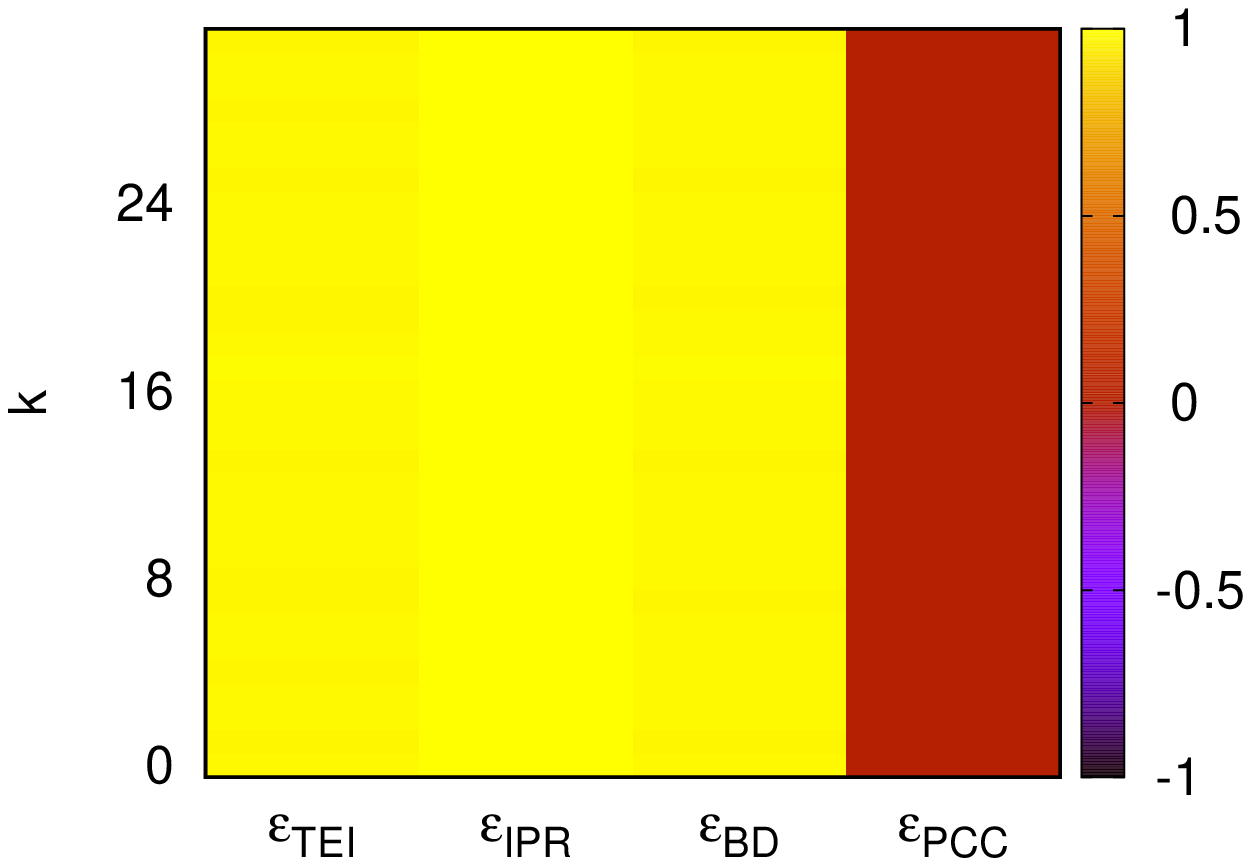}
\caption{Correlation of $\xi_{\textsc{svne}}$ 
with 
 $\xi$-indicators (left) and   
 with $\neweps$-indicators for the slice 
 corresponding to $\theta=\frac{1}{2}\pi$ for the field and 
 the $\sigma_{x}$ basis for each qubit (right).
 The figures are 
 for the eigenstates $\ket{\psi_{5,6,k}}, 
 \,0\leqslant k \leqslant 2^{5}-1$ in Case (i)  in the Tavis-Cummings model.
}
\label{fig:CorrPlotNat5Ntc6TCMVaryg}
\end{figure}

 We  have considered three cases, namely,
(i)\, $\Lambda_{s} = \chi = 0$ \,(ii)\,   
$\Lambda_{s}/(2 \pi) = 1\,{\rm MHz}, \, \chi=0$ 
\, (iii)\, $\Lambda_{s}/(2\pi) = 
\chi/(2\pi)  = 1\, {\rm  MHz}$.
In each case, $\Lambda/(2\pi) $ is varied from 
$- 1.2\,{\rm  MHz}$  to $ 1.3\,{\rm  MHz}$  
in steps of $0.025\,{\rm  MHz}$. It is easily 
shown that the total number operator 
\begin{equation}
\mathcal{N}_{\mathrm{tot}}
=a^{\dagger}a + \sum_{p=1}^{M} \sigma_{p}^{+} \sigma_{p}^{-},
\label{eq:ntotTC}
\end{equation} 
commutes with  $H_{\textsc{tc}}$.  For each value of $\Lambda$ we have numerically solved for the  complete set 
$\lbrace \ket{\psi_{M,N,k}} \rbrace$ 
of common eigenstates of 
 $\mathcal{N}_{\mathrm{tot}}$ and 
 $H_{\textsc{tc}}$, 
where $N = 0,1,\ldots$ is the eigenvalue of 
 $\mathcal{N}_{\mathrm{tot}}$ and 
 $k = 0, 1, \ldots, 2^{M}-1$. 
   Considering the  total system as a bipartite composition of the field subsystem and a subsystem comprising all 
 the qubits, we have computed the entanglement 
 indicators. Figure \ref{fig:CorrPlotNat5Ntc6TCMVaryg} 
 shows   the correlation between the indicators  and 
 $\xi_{\textsc{svne}}$ in Case (i). The associated  
 Pearson correlation coefficients are $0.97$ for 
$\epsarg{tei}$,  $0.99$ for $\epsarg{ipr}$,
$0.97$ for $\epsarg{bd}$, correct to 
two decimal places. 
(The accuracy  of the  $\neweps$-indicators depends, of course,  on the basis chosen.) 
On averaging, we obtain 
the corresponding $\xi$-indicators  
with a PCC equal to $0.99$, showing that these indicators   track  $\xi_{\textsc{svne}}$ very closely. 
We have carried out a similar exercise in Cases 
(ii) and (iii). The results and the inferences drawn from them 
are broadly similar to those 
found in Case (i).  

Finally, with  $\Lambda/(2 \pi)$ set equal to  
 $1.2 \,{\rm  MHz}$, we have examined the effect of changing 
 the  strength of the disorder in $\Omega_{p}$ 
  by varying the standard deviation 
  of $\Delta_{p}$ 
   from $0$ to $0.2 \,\aver{\Delta}$ in steps of 
   $2\times 10^{-4}\, \aver{\Delta}$. Calculating the entanglement indicators for each disorder strength in 
   $\Omega_{p}$, we have found the correlations between the $\xi$-indicators and $\xi_{\textsc{svne}}$ in 
   Cases (i), (ii), and (iii). 
     $\xi_{\textsc{tei}}$ and $\xi_{\textsc{bd}}$ turn out to be significantly closer to 
  $\xi_{\textsc{svne}}$, and hence more accurate 
  indicators of entanglement, 
    than the other indicators.

\section{\label{sec:conc}Concluding remarks}
We have considered  generic bipartite continuous-variable systems and hybrid quantum systems in the presence of nonlinearities, and  
tested quantitatively the efficacy of various indicators in estimating entanglement directly from quantum state tomograms close to avoided 
energy-level crossings. We find that the nonlinear correlation between the respective quadratures 
of the  two subsystems reflects very reliably  the extent of entanglement in bipartite CV systems governed by number-conserving Hamiltonians.  We have shown that if the eigenstates of the Hamiltonian are Hamming-uncorrelated, the inverse-participation-ratio-based 
quantifier  $\xi_{\textsc{ipr}}$ is an  excellent indicator of entanglement near avoided  crossings. In fact, even $\epsarg{ipr}$ (the corresponding indicator for a single section of the tomogram) suffices to estimate entanglement reliably. 
The tomographic entanglement indicator 
$\xi_{\textsc{tei}}$ and the Bhattacharyya-distance-based indicator $\xi_{\textsc{bd}}$ are  also good indicators  at avoided crossings, in contrast to the linear correlator $\xi_{\textsc{pcc}}$ which is based on the Pearson correlation coefficient.  
 Entanglement indicators seem to perform better with increasing   
$\aver{N_{\mathrm{tot}}}$.  
The conclusions drawn are both significant and readily applicable in identifying optimal entanglement indicators that are easily obtained from tomograms, without employing state-reconstruction procedures.
\section*{References}
\bibliography{references}

\end{document}